\documentclass[nofootinbib,a4paper,10pt,twocolumn,superscriptaddress]{revtex4-2} %

\usepackage{amsmath,amssymb,graphicx,mathrsfs}
\usepackage{appendix,bm,dcolumn}
\usepackage[T1]{fontenc}
\usepackage{color}
\usepackage[unicode=true,pdfusetitle,bookmarks=true,bookmarksnumbered=false,bookmarksopen=false,breaklinks=true,pdfborder={0 0 0},backref=false,colorlinks=true]{hyperref}
\hypersetup{citecolor=blue,filecolor=blue,linkcolor=blue,urlcolor=blue}
\usepackage{aas_macros} 

\usepackage[perpage]{footmisc}

\newcommand{\lcdm}{$\Lambda {\rm CDM}$}
\newcommand{\flrwsolver}{\texttt{FLRWSolver}}

\newcommand{\alp}{\alpha}

\begin{document}

\title[Lensing convergence in general relativity]{Large-scale weak lensing convergence in nonlinear general relativity}

\author{Hayley J. Macpherson}
\address{Kavli Institute for Cosmological Physics, The University of Chicago, 5640 South Ellis Avenue, Chicago, Illinois 60637, USA}
\email{hayleyjmacpherson@gmail.com}
\vspace{10pt}
\date{\today}

\date{\today}%

\begin{abstract}
In this work we investigate the weak lensing convergence using an end-to-end nonlinear general relativistic framework. Combining numerical relativity simulations of large-scale structure formation with general relativistic ray-tracing, we compare our nonlinear calculation to the expectation based on perturbation theory for a set of 20 synthetic observers.
We focus on large angular scales $\ell < 100$ across a broad range of redshifts with $0.05<z<3$. We confirm the importance of Doppler lensing for redshifts below $z\sim$~0.6, as predicted by previous works. On average across our observers, linear perturbation theory predicts the nonlinear convergence to within 3--30\% across all redshifts and angular scales we study. 
In general, we find smaller angular scales are better matched by linear theory than larger angular scales. 
While we cannot definitively identify the source of the discrepancy, for our particular study of redshift slices on observers' light cones the differences are mostly below the level of cosmic variance.
\end{abstract}

\maketitle

\section{Introduction}

Weak lensing of galaxy images---as well as the cosmic microwave background (CMB)---are powerful probes %
of the standard
$\Lambda$ Cold Dark Matter (\lcdm) cosmological model. 
\lcdm\, at its core assumes general relativity (GR) as well as the exactly spatially homogeneous and isotropic Friedmann-Lema\^itre-Robertson-Walker (FLRW) class of models to describe the large-scale dynamics of the Universe.
This relatively simple model has enjoyed over two decade of success in its ability to consistently fit most of our cosmological observations. However, as our data become more precise we are beginning to see some cracks forming in \lcdm\, \citep[see][for a review]{Perivolaropoulos:2022}. These include the long-debated `Hubble tension', the $S_8$ tension \citep[though see][]{Wright:2025}, the large-scale CMB anomalies, and the cosmic dipole tension. More recently, there is increasing excitement surrounding the potential dynamic nature of dark energy since the release of the latest results from the Dark Energy Spectroscopic Instrument \citep[DESI;][]{DESI-Collab-2025}. 

The validity of the FLRW models in approximating the late-time Universe has long been called into question \citep[see, e.g.][for reviews and discussions]{Bolejko:2017,Coley:2020,Kumar-Aluri:2023,Akarsu:2024}.
While the presence of inhomogeneities necessitates that the Universe cannot be \textit{exactly} described by an FLRW model on any scale \citep[e.g.][]{Rasanen:2011,Buchert:2012,Green:2014,Buchert:2015}, it may be close enough that the difference falls within the uncertainties in our measurements.
Importantly, in recent years general-relativistic simulations indicate that the size of this so-called `backreaction' effect is indeed very small %
\citep{Bentivegna:2016a,Adamek:2019,Adamek:2019b,Macpherson:2019,Macpherson2023}.

As well as choosing a background cosmology, %
we also need to %
model the inhomogeneities in the Universe that give rise to structure formation. 
Which set of approximations are adopted in modeling these inhomogeneities typically depends on the particular regime we are interested in. For the case of weak lensing of galaxy images, it is typically assumed that linear theory is sufficient to describe the effect\footnote{In the particular case of the lensing convergence we will study here, some linear terms are also neglected. See Section~\ref{sec:kappa_approx}.}. 
The scalar perturbation to the metric---the Newtonian gravitational potential---is widely known to remain small on scales above those of compact objects. However, 
\textit{derivatives} of the potential (e.g. density contrasts) can be large in the late Universe \citep[which are relevant for observables such as distances;][]{Bonvin:2006}.
This can invalidate results obtained in linear perturbation theory that assume \textit{all} perturbations are small \citep{Rasanen:2010}.
Most methods for approximating inhomogeneities, including in lensing, have not been directly tested against non-perturbative solutions in GR. In the now-imminent era of precision cosmology, it is vital to ensure that the accuracy of our approximations is better than the \textit{precision} of upcoming observations from, e.g., \textit{Euclid} \citep{Deshpande:2020} or the Vera Rubin observatory's Legacy Survey of Space and Time \citep[LSST;][]{LSSTScienceBook:2009}.

One of the best ways we can test our approximations is to remove them from our cosmological modeling and see if they hold. 
Numerical relativity (NR) is a computational method to evolve Einstein's equations of GR without imposing space-time symmetries; it is a vital component of the observation of gravitational waves from binary compact objects \citep{Lehner:2014}. NR has more recently proven to be a useful tool for the study of late-time cosmology
\citep[e.g.][]{giblin:2016a,Bentivegna:2016a,Macpherson:2017}.
Especially, combining general-relativistic ray tracing with NR simulations means we can now study observations of a nonlinear cosmic web in a simulation which is free from common gravitational and geometric approximations \citep{giblin2016b,east:2018a,Tian:2021a,Macpherson:2021,Macpherson2023}.

\citet{Giblin:2017a} presented the first study of the weak lensing convergence signal in a cosmological simulation with NR. %
The authors compared their fully nonlinear GR convergence to the widely-adopted approximation derived from perturbation theory, finding a $\sim$2--20\% difference in the angular power spectra (depending on the angular scale).
However, this work focused only on very large angular scales ($\ell\sim 10$) and a single redshift of $z=0.25$. 
In this paper, we aim to extend the work of \citet{Giblin:2017a} to a more realistic simulated scenario. Specifically, we will use simulations with redshift zero density fluctuations in line with observations, study 
multiple observers in different environments, extend to slightly smaller angular scales, and sample a wide range of redshifts up to $z=3$.

In Section~\ref{sec:NRsims} we introduce our NR simulations, in Section~\ref{sec:rt} we discuss the post-processing framework we use to generate synthetic observables; including the lensing convergence both from ray tracing and from perturbation theory. 
In Section~\ref{sec:results} we present and discuss our results and we conclude in Section~\ref{sec:conclude}. Our Appendices show the numerical robustness of our results (including convergence tests and constraint violation), study the Born approximation, and (unsuccessfully) investigate potential causes of the difference we find. We only display results in the main text that are robust to changes in numerical resolution. 
We adopt geometric units with $c=G=1$, Latin letters represent spatial indices and take values $1\ldots 3$, Greek letters represent space-time indices and take values $0\ldots 3$, and we adopt the Einstein summation convention with repeated indices implying summation.

\section{Numerical relativity simulations}\label{sec:NRsims}

The simulations we use were performed with the Einstein Toolkit %
\citep[ET;][]{Loffler:2012,Zilhao:2013}; an open-source NR code based on the Cactus\footnote{\url{https://www.cactuscode.org}} 
infrastructure. We adopt the Baumgarte--Shapiro--Shibata--Nakamura \citep[BSSN;][]{Baumgarte1998,Shibata1995} formalism, which makes use of a 3+1 decomposition of space-time. In this decomposition, spatial hypersurfaces are defined by their time-like normal vector $n^\mu$ which is defined by arbitrary choices of the lapse function, $\alpha$, and the shift vector, $\beta^i$. This decomposition splits Einstein's equations into a set of evolution equations for the spatial metric tensor, $\gamma_{ij}$, and extrinsic curvature, $K_{ij}$, as well as two constraint equations: the Hamiltonian and momentum constraints,
\begin{subequations}\label{eqs:constraints}
\begin{align}
    H &\equiv \mathcal{R} - K_{ij}K^{ij} + K^2 - 16\pi\rho = 0, \\
    M_i &\equiv D_j K^j_{\phantom{j}i} - D_i K - 8\pi S_i = 0,
\end{align}
\end{subequations}
respectively. Here, $\mathcal{R}$ is the spatial Ricci scalar, $K=\gamma^{ij}K_{ij}$ is the trace of the extrinsic curvature, $\rho\equiv T_{\mu\nu} n^\mu n^\nu$ is the energy density projected into the hypersurface frame (with $T_{\mu\nu}$ the stress-energy tensor), $D_i$ is a covariant deriviative associated with $\gamma_{ij}$, and $S_i=\gamma_{i\alp}n_\beta T^{\alp\beta}$ is the momentum density of the hypersurfaces. 
The set of equations \eqref{eqs:constraints} are identically zero for exact solutions of Einstein's equations and so are typically used in NR to set initial data as well as to keep track of the accumulation of numerical finite-difference errors during the simulation. 

In the ET, we use the \texttt{McLachlan} thorn \texttt{ML\_BSSN} \citep{Brown:2009} to evolve the space-time using a harmonic-like gauge with zero shift vector \citep[see][for specifics about the gauge]{Macpherson:2019}. The hydrodynamics is evolved with the \texttt{GRHydro} \citep{Baiotti:2005} thorn using a polytropic equation of state with $P\ll\rho$ via \texttt{EOS\_Omni}. This setup has been shown to accurately reproduce a perturbed Einstein-de Sitter (EdS) model evolution \citep{Macpherson:2017}. Especially, the use of a polytropic equation of state in place of a pure dust ($P=0$) evolution has been suggested as a useful alternative for inhomogeneous fluid simulations to avoid coordinate collapse in the non-linear regime \citep{BolejkoLasky2008}.

\subsection{Initial data}\label{sec:ics}

Our simulations were initialised using the ET thorn \flrwsolver\, \citep{Macpherson:2017}, which 
sets the initial metric tensor $\gamma_{ij}$ as a perturbed EdS space-time with 
\begin{equation}\label{eq:pertmetric}
    ds^2 = - a^2 (1 + 2 \phi)\, dt^2 + a^2(1-2\phi)\,\delta_{ij}\, dx^i \, dx^j,
\end{equation}
where $a=a(t)$ is the background scale factor, $\phi$ is the scalar gravitational potential, $x^\mu = (t, x^i)$ are the simulation coordinates, and $\delta_{ij}$ is the Kronecker delta function. 

\flrwsolver\, allows for several forms of $\phi$ to define \eqref{eq:pertmetric} \citep[and the resulting matter perturbations; see, e.g.][]{Macpherson:2017,Macpherson:2019}. In this work, we adopt the new `exact' perturbation as first introduced in Section~3.2 of \citet{Macpherson:2025aa}. In this method, the initial power spectrum $P_\phi(k)$ is supplied and \flrwsolver\, generates $\phi(x^i)$ as a Gaussian random field following this power spectrum. This field is input into the spatial metric $\gamma_{ij}$ via \eqref{eq:pertmetric} and $K_{ij}$ via the 3+1 relation $-2\alpha K_{ij} = \partial_t \gamma_{ij}$ (for zero shift).  \flrwsolver\, then calculates the geometric terms on the left hand side of the Hamiltonian and momentum constraints \eqref{eqs:constraints}, finally inverting these equations to obtain the density and velocity (the latter via the momentum density $S_i$) \citep[see also][]{Adamek:2020}. 
This method of initial data generation offers an improvement over earlier work using \flrwsolver\, which made use of linear perturbation theory on the initial slice \citep[e.g.][]{Macpherson:2017,Macpherson:2018,Macpherson:2019,Macpherson:2021,Heinesen:2022,Macpherson2023,Koksbang:2024,Williams:2024}. In these works, the constraints \eqref{eqs:constraints} were thus solved only to \textit{linear order} in all perturbations and their derivatives. With this new method the constraints are solved `exactly' (i.e., to the numerical precision of the finite-difference derivatives used to calculate $\mathcal{R}$ and derivatives of $K_{ij}$) and thus the initial data is correct to all orders in the perturbations given the metric \eqref{eq:pertmetric}. This allows us to begin the simulations at lower redshift (namely, $z=20$ rather than $z=1000$ as in previous works) which reduces the accumulation of constraint violation by the time the simulation has reached $z\approx 0$ by an order of magnitude. 

We emphasize that the form of the metric \eqref{eq:pertmetric} is enforced on the initial slice only and the evolution of the simulations follows the generic BSSN framework without any `background' cosmology or perturbative assumptions. 

\subsection{Simulation specifics for this work}\label{sec:sims}

To define our simulations we need to specify only the initial metric fluctuations via $P_\phi(k)$ and define the initial background cosmology.
Briefly, we make these choices such that the density field at $z=0$ in the simulation matches observed qualities of the Universe as closely as possible. However, we note the evolution of the simulations \textit{do not contain a cosmological constant}, $\Lambda$\footnote{This is simply because there is no inclusion of $\Lambda$ in the evolution of the Einstein equations in the ET; this development is in progress and we plan to include $\Lambda$ in future work.}. We also note that our simulations do not contain particles and instead adopt a continuous fluid approximation for the evolution of the dark matter \citep[though see][for the preliminary inclusion of particles in the ET]{Magnall:2023}.

In more detail, we generate $P_\phi(k)$ at $z=0$ via the CAMB code \citep{Lewis2000} with $\Lambda$CDM parameters $A_s=2\times 10^{-9}$, $k_{\rm pivot}=0.05 \,h$/Mpc, $n_s=0.965$, Hubble constant $h=0.7$ (where $H_0=100\,h$ km/s/Mpc), baryonic and CDM energy content $\Omega_b=0.049$ and $\Omega_{\rm cdm} = 0.25-\Omega_b$, curvature $\Omega_k=0$, and tensor-to-scalar ratio $r=0$. To obtain the potential power spectrum $P_\phi$, we convert from the Weyl potential ($\Psi$; as output in CAMB) via $P_\Psi = k^4 P_\phi$. %
Since the evolution of our simulations is matter-dominated, we have $\partial_t \phi=0$ (in the linear regime) and so we set the power spectrum at $z=0$ to be the same as for our initial data at $z_{\rm ini}=20$. This also enters in our calculation of $K_{ij}$ from $\gamma_{ij}$ which also assumes $\partial_t\phi=0$. 

We define the initial background cosmology as an EdS model with parameters $\Omega_{\rm cdm}=1-\Omega_b$ (with $\Omega_b$ as above) and $h=0.7$. We use the CAMB Python wrapper to calculate the initial Hubble parameter $H_{\rm ini}$, the initial coordinate time is $t_{\rm ini}=2/H_{\rm ini}$ \citep[since in our gauge $t$ coincides with conformal time on the initial slice, see][]{Macpherson:2019}, and the initial scale factor is $a_{\rm ini}=1/(1+z_{\rm ini})$. Note that the cosmology used to generate $P_\phi$ is different to the one used to scale it back to the initial data, and for the initial background itself. This method yields a power spectrum of density fluctuations on the $z=0$ simulation slice which roughly matches that of the observed $\Lambda$CDM power spectrum at $z=0$, which we confirm once the simulation is finished. 

Our fiducial simulation has co-moving cubic domain length of $L=3072 \,h^{-1}$ Mpc and numerical resolution of $N=256$ grid cells on each side. All results presented in the main text are calculated in this simulation. 
To study the numerical convergence of our results, we also perform two lower-resolution simulations with $N=128$ and $N=200$ which sample the same physical scales as the fiducial simulation. The initial data for these simulations is generated from the same $P_\phi(k)$ as the high-resolution simulation with a different random seed. 
With these, we ensure all our results are unchanging with numerical resolution and only present results in the main text that we consider to be converged in this sense (see \ref{appx:convergence} for more details). All simulations adopt periodic boundary conditions throughout the evolution.

When simulating a power spectrum of fluctuations on an Eulerian grid, modes with wavelengths close to the grid scale are not properly resolved. Including such modes can lead to an unsatisfactory amount of constraint violation sourced by numerical error. To minimise such errors, we remove modes from the initial power spectrum with wavelengths below a chosen scale for all simulations. 
Specifically, we set $P_\phi(k>k_{\rm cut})=0$ where $k_{\rm cut}=\sqrt{k_{x,{\rm max}}^2 + k_{y,{\rm max}}^2 + k_{z,{\rm max}}^2}$ and $k_{x,{\rm max}}=k_{y,{\rm max}}=k_{z,{\rm max}}=2\pi/\lambda_{\rm min}$. We set $\lambda_{\rm min}=10 \,\Delta x$, where $\Delta x$ is the grid spacing for our fiducial simulation with $N=256$. For the lower resolution simulations, the same physical scale corresponds to $\lambda_{\rm min}=5\,\Delta x$ (for $N=128$) and $\lambda_{\rm min}=7.8125\,\Delta x$ (for $N=200$). This corresponds to a physical scale of $\lambda_{\rm min} \sim 120\,h^{-1}$~Mpc.
We note that this scale cut is only enforced for the initial data, and structure below this scale will (and does) grow as the simulation becomes nonlinear, however, the power at these scales is significantly damped with respect to a \lcdm\, power spectrum. 

While this scale cut is necessary to ensure our results are robust to numerical resolution, we expect that reducing $\lambda_{\rm min}$ would affect our results since the physical problem is changing. Our fiducial simulation has grid cells of $12\,h^{-1}$Mpc in size, which is approximately the size of galaxy clusters. We are currently limited to grid cells of this size since the continuous fluid approximation is only valid above the scales of bound objects. To study smaller physical scales would require the inclusion of particles with the ET \citep{Magnall:2023}.

\section{Ray-tracing and post-processing analysis}\label{sec:rt}

We now want to calculate the convergence signal for observers who live in the simulated universe. 
To this end, we %
use the \texttt{mescaline} ray tracer \citep{Macpherson2023} in post processing. For a specified observer position and directions of observation, \texttt{mescaline} solves the geodesic deviation equation to advance ray bundles along geodesics through the inhomogeneous space-time output from the simulation. Observers and `sources' are both co-moving with the fluid flow, specified by the value of the fluid 4--velocity $u^\mu$ at their space-time position. In practice this means that the photon energy is calculated as $E\equiv -k^\mu u_\mu$, where $k^\mu$ is the photon 4--momentum. The redshift, $z$, of a source, $s$, as measured by an observer, $o$, is defined as the ratio of energies $1+z\equiv E_s/E_o$. The angular diameter distance is calculated by directly propagating the Jacobi matrix along geodesics; sourced by the Ricci and Weyl lensing scalars. See \citet{Macpherson2023} for complete details and tests of the \texttt{mescaline} ray tracer.

In this work, we randomly position 20 observers in each of the three simulations introduced in Section~\ref{sec:sims}. 
Individual rays for each observer are initially directed in the center of \texttt{HEALPix}\footnote{\url{https://healpix.sourceforge.io}} pixels with $N_{\rm side}=32$. We note that this $N_{\rm side}$ is used in the ray tracer \textit{only} to determine the incoming direction of the rays to the observer and the code follows the fully perturbed path of each geodesic. However, it defines the resolution of the output maps and thus the resulting angular power spectra.

Each ray bundle is propagated to a redshift approximately coincident with one light-crossing distance of the simulation box. Since we are propagating a full sky of rays, we stop the analysis here to avoid multiple samplings of the same structures in the box. The periodic boundary conditions employed in the simulation are used in the ray tracer when geodesics cross the boundary.
Our simulations have box size $L=3072\,h^{-1}$ Mpc and thus the maximum redshift is $z_{\rm max}\approx 3$ (for an EdS universe). We also make a minimum redshift cut of $z\approx0.05$ for all analyses that follow. 

It has been shown that distances and redshifts calculated from NR simulations using \texttt{mescaline} are accurate to within $10^{-5}$ and $5\times 10^{-4}$, respectively \citep[see Appendix E4 of][]{Macpherson2023}. Currently only the spatial averaging part of \texttt{mescaline} is public\footnote{\url{https://github.com/hayleyjm/mescaline-1.0}} \citep[as used in][]{Macpherson:2019}, and we intend to include the ray tracer in the near future.

\subsection{Nonlinear lensing convergence}

The Jacobi matrix encodes all of the distortions of an infinitesimal ray bundle, as seen by an observer, as it traverses a generic inhomogeneous space-time. The trace of the Jacobi matrix is the angular diameter distance, $D_A$, and its four components can be decomposed to describe the shear and rotation of images \citep[see Section~2.2 of][]{Fleury:2015a}. This decomposition can be done without prescribing anything about the space-time distorting the image.

In the case where we have some perturbed background cosmology---and thus the shear and rotation are very small---we can additionally define the \textit{lensing convergence} $\kappa$. This is defined as a perturbation of the angular diameter distance,
\begin{equation}\label{eq:kappadef}
    \kappa \equiv \frac{\bar{D}_A - D_A}{\bar{D}_A},
\end{equation}
with respect to the background distance $\bar{D}_A$. 
Since our NR simulations do not adopt a perturbative description in the evolution, we must \textit{assume} the existence of a well-defined background cosmology. 
In this work, we choose the background EdS model which we used for our initial data, and thus $\bar{D}_A = D_{A,{\rm EdS}}(z_s)$ where $z_s$ is the \textit{observed} redshift of each source\footnote{This is distinct from the redshift of the background model at the time coordinate associated with the light-cone slice; these differ by up to $\sim 10\%$ at low redshift} and thus is different for each line of sight.

In previous works, we found that the large-scale spatial average over simulation hypersurfaces matched very closely to this initial model \citep[i.e., there is minimal backreaction on spatial hypersurfaces of the simulations;][]{Macpherson:2018,Macpherson:2019}. This conclusion of course depends on the specific spatial slicing chosen \citep{Adamek:2019,Giblin:2019}, and the \textit{observational} impact of inhomogeneities on FLRW parameters in the context of NR simulations can be larger \citep{Macpherson:2024} (and \citep{Elgamal_inprep}; in prep.).

The \texttt{mescaline} ray tracer calculates and outputs the Jacobi matrix in all generality from the simulated space-time. This yields angular diameter distances and redshifts along each geodesic spaced on the constant-coordinate time slices of the simulation output. In other words, the light cone for each observer is sliced in spheres of constant simulation coordinate time. We use this data to define the convergence via \eqref{eq:kappadef}. %

We denote the calculation of the convergence via \eqref{eq:kappadef} the \textit{`nonlinear convergence'} since it is the most generic way to define the signal and in principle contains perturbative contributions at all orders. In the next section, we will introduce what we call the \textit{`linearised convergence'} which is the description commonly adopted in standard cosmology which is limited only to linear terms (with some first-order terms also neglected).

\begin{figure*}%
    \centering
    \includegraphics[width=\linewidth]{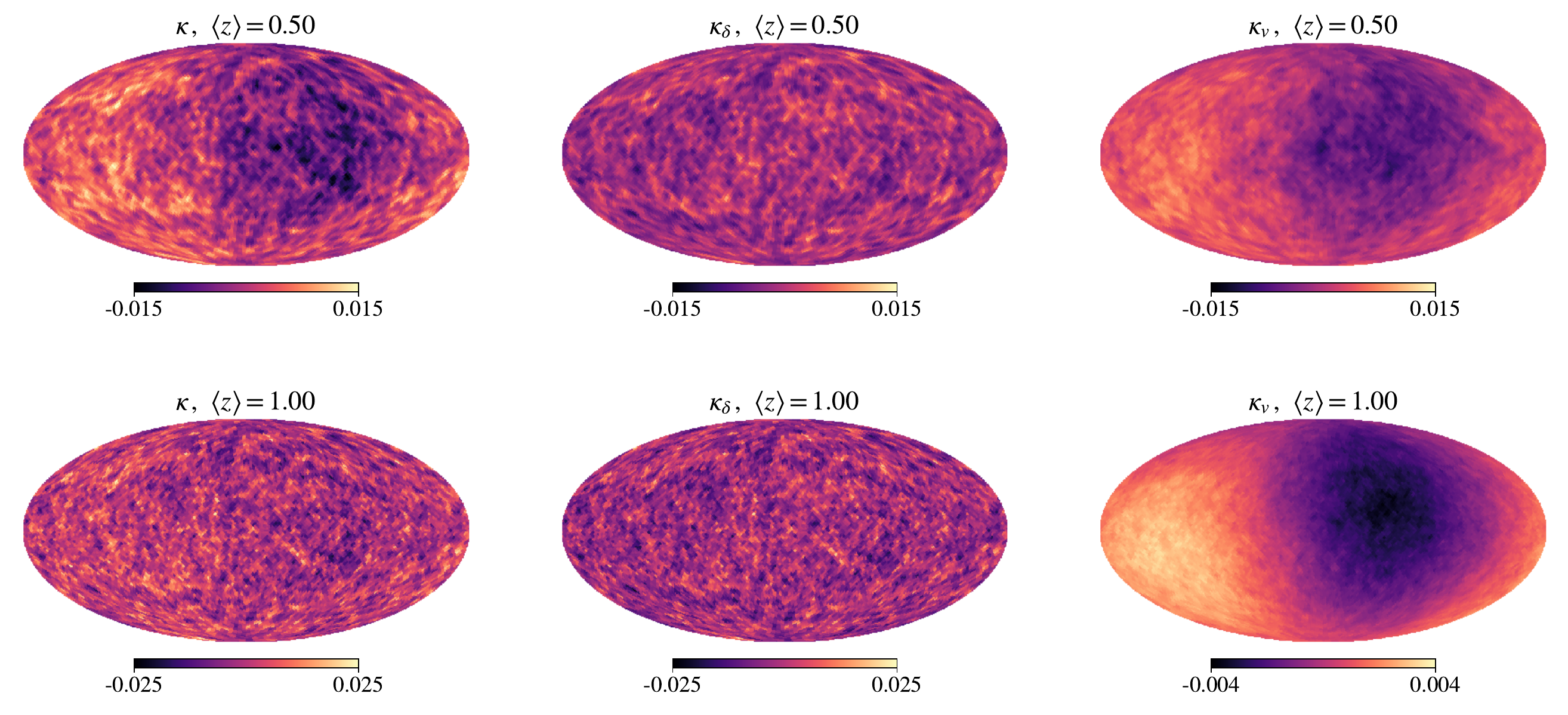}
    \caption{Full sky maps of the nonlinear convergence (left columns), the linearised convergence from the density field (center columns), and the Doppler lensing (right columns) for three slices with mean redshifts $\langle z \rangle=0.5$ (top row) and $\langle z \rangle=1.0$ (bottom row). These maps are for one single observer in the simulation with $N=256$. Note the colorbar limits differ for the bottom right panel since the Doppler lensing signal is much smaller at higher redshifts.}
    \label{fig:skymaps}
\end{figure*}

\subsection{Linearised lensing convergence}\label{sec:kappa_approx}

We wish to compare the nonlinear convergence defined in the previous section to the common approximation used in weak lensing analyses. Typically, these make use of the linearly-perturbed FLRW model with metric defined in \eqref{eq:pertmetric}. Combined with this metric, we can derive the following form of the convergence from the general definition \eqref{eq:kappadef} \citep{Das2008,Fleury:2015a}
\begin{equation}\label{eq:kappaapproxdef}
    \kappa_{\delta} \equiv \frac{3}{2}H_0^2 \Omega_m \int^{\chi_s}_0 \frac{\chi (\chi_s - \chi)}{\chi_s} \frac{\delta(\chi)}{a(\chi)} {\rm d}\chi. 
\end{equation}
Here, $\Omega_m$ %
is the cosmological density parameter for the background FLRW model, 
$H_0\equiv H(z=0)$ is the proper-time Hubble parameter at redshift zero (or Hubble constant), 
$\chi_s$ is the background co-moving distance to the source at which $\kappa_\delta$ is measured, 
$\delta\equiv\rho_0/\bar{\rho}_0-1$ 
is the dimensionless density contrast at $\chi$ 
(with $\rho_0\equiv T_{\mu\nu}u^\mu u^\nu$ the rest-mass density), and $a(\chi)$ is the background FLRW scale factor. In Appendix~\ref{appx:rel_calcs} we provide details on exactly how we calculate $\kappa_\delta$ from our simulation data.

In deriving \eqref{eq:kappaapproxdef}, all second-order contributions have been neglected {as well as} some first-order contributions. Specifically, in deriving the perturbed Jacobi map several terms of $\mathcal{O}(\phi)$, $\mathcal{O}(\partial_i \phi)$, and $\mathcal{O}(\partial^2_\chi \phi)$ are neglected because they are expected to be sub-dominant to second spatial derivatives of $\mathcal{O}(\partial^2 \phi)$. Further, no perturbation in the frequency of the light ray has been considered in deriving this expression---leading to Doppler contributions that can be large at low redshifts \citep[e.g.][]{Bonvin:2008,Bolejko:2013,Bacon:2014,Fleury:2017}. %
After neglecting these terms, the remaining contribution can then be simply related to an integral of the density field along the line of sight via the Poisson equation, finally leading to \eqref{eq:kappaapproxdef}. 
A full derivation of \eqref{eq:kappaapproxdef} in the perturbation theory context---starting from the fully nonlinear relation \eqref{eq:kappadef}---can be found in Section~5.2.2 of \citet{Fleury:2015a}. We explore the contribution of these neglected terms in our simulations in Appendix~\ref{appx:kappa_terms}. 

Commonly, \eqref{eq:kappaapproxdef} is calculated by adopting the Born approximation: neglecting the perturbations to the photon path as it traverses the inhomogeneous space-time and treating it as a straight line. Such an approximation has been shown to be excellent in Newtonian $N$-body simulations \citep{Ferlito:2024,Fabbian:2018,Hilbert:2009} and perturbation theory \citep{Pratten:2016}. 
For our fiducial results, our calculation of $\kappa_\delta$ includes $\delta(\chi)$ calculated along fully perturbed paths (albeit with $\chi$ calculated via the EdS relation). 
In Appendix~\ref{appx:chi_Born} we explore the validity of the Born approximation in our simulations and find it holds at the $\sim 0.1\%$ level; broadly consistent the aforementioned studies.%

\subsubsection{Known general-relativistic corrections}\label{sec:gr_kappas_approx}

It is known that the contribution from the density \eqref{eq:kappaapproxdef} does not give the whole picture in all cases.
General-relativistic contributions to the lensing convergence have been studied in the context of perturbation theory or post-Newtonian approximations \citep[e.g.][]{Bonvin:2006,Bonvin:2008,Andrianomena:2014,Umeh:2014,Thomas:2015b}, exact solutions \citep[e.g.][]{Bolejko:2013,Troxel:2014,Koksbang:2019} and simulations \citep[e.g.][]{Giblin:2017a,Lepori:2020,Grimm:2021,Coates:2021,Rasera:2022}. Notably, the Doppler contribution has been shown to be dominant on large angular scales and low redshifts \citep{Bacon:2014,Fleury:2017} which are relevant for this work.

In perturbation theory, the total linearised convergence can be decomposed as
\begin{equation}\label{eq:ktot}
    \kappa_{\rm tot} = \kappa_\delta + \kappa_v + \kappa_{\rm SW} + \kappa_{\rm ISW},
\end{equation}
where $\kappa_\delta$ is defined above in \eqref{eq:kappaapproxdef}, and the other contributions
\begin{subequations}\label{eqs:grkappas}
    \begin{align} 
        \kappa_v &= \frac{1+z_s}{H \chi_s} {\bf v_o} \cdot {\bf n} + \left(1 - \frac{1+z_s}{H \chi_s} \right){\bf v_s} \cdot {\bf n}, \label{eq:kappadop} \\
        \kappa_{\rm SW} &= 2\,\phi_s - \phi_o + \frac{1+z_s}{H \chi_s} \left( \phi_o - \phi_s \right), \label{eq:kappasw}\\
        \kappa_{\rm ISW} &= -\frac{2}{\chi_s} \int^{\chi_s}_{0} \phi\, {\rm d}\chi + 2 \left(1 - \frac{1+z_s}{H \chi_s} \right) \int_{0}^{\chi_s} \phi'\, {\rm d}\chi, \label{eq:kappaisw}
    \end{align}
\end{subequations}
are the Doppler, Sachs-Wolfe (SW), and Integrated Sachs-Wolfe (ISW) effects, respectively \citep[see, e.g.][for a more complete derivation]{Bacon:2014}. Note that $\kappa_v$ is sometimes referred to as the `Doppler lensing', although this term merely captures the effects of peculiar velocities on the observables. Throughout this paper, we will use $\kappa_{\rm tot}$ to denote the `total' linearised convergence when considering different combinations of contributions on the right hand side of \eqref{eq:ktot}. 
In \eqref{eqs:grkappas} above, subscripts $s$ and $o$ imply that quantity is evaluated at the space-time position of the source and observer, respectively, a prime is a derivative with respect to conformal time, i.e. $'\equiv \partial_\eta$. The potential is the one in the perturbed metric \eqref{eq:pertmetric}, the peculiar velocity is ${\bf v} = v^i = dx^i/dt$, and ${\bf n}$ is the direction vector pointing from source to observer (i.e., $-{\bf n}$ is the direction of observation). %

The convergence sourced by the density fluctuations, \eqref{eq:kappaapproxdef}, is expected to be the far dominant contribution for cosmological constraints and thus  
is widely adopted in galaxy \citep{DES-lensing-2022} and CMB \citep{Planck-lensing-2020,Lewis:2006} lensing analyses. 
The Doppler contribution \eqref{eq:kappadop} is dominant for low redshifts and large angular scales and could be important for wide-sky surveys such as \textit{Euclid} \citep[see, e.g.][]{Grimm:2021}. The SW contributions \eqref{eq:kappasw} and \eqref{eq:kappaisw} are typically neglected and considered sub-dominant due to the relative small amplitude of $\phi$ with respect to $\delta$ and ${\bf v}$.

In this work, we will extend previous analyses of the relativistic contributions \eqref{eqs:grkappas} and calculate each signal in NR simulations for the first time \citep[with the exception of the Doppler contribution which was also studied in][]{Giblin:2017a}. We will compare individual contributions, as well as their sum, to the nonlinear convergence \eqref{eq:kappadef}. See Appendix~\ref{appx:rel_calcs} for specific details on how we extract the potentials and velocities in \eqref{eqs:grkappas} from our NR simulation data.

\section{Results \& Discussion}\label{sec:results}

He we describe the results of comparing the nonlinear convergence $\kappa$ in \eqref{eq:kappadef} to the linearised contributions \eqref{eq:kappaapproxdef} and \eqref{eqs:grkappas} for the same set of observers. In Section~\ref{sec:results_skymaps} we assess the full-sky distribution of the convergence signals, in Section~\ref{sec:results_Cls} and \ref{sec:Clzevol} we study the angular power spectra, and in Section~\ref{sec:LOS} we briefly assess individual lines of sight. 

\subsection{Sky maps}\label{sec:results_skymaps}

Figure~\ref{fig:skymaps} shows a set of sky maps (in Mollweide projection) of the nonlinear convergence $\kappa$ (left columns), the linearised convergence from the density field $\kappa_\delta$ (center columns), and the Doppler lensing $k_v$ (right columns) for one observer in the $N=256$ simulation. The redshift slices are $\langle z \rangle = 0.5$ for the top row and $\langle z \rangle = 1.0$ for the bottom row, where $\langle\rangle$ represents an average over the observer's sky. 
For $\langle z \rangle = 0.5$, we can see by eye the difference between $\kappa$ and $\kappa_\delta$, but we can also see that the dipole signature visible in $\kappa$ appears to be well-represented by $\kappa_v$ in the right-most column for this particular observer. At higher redshift of $\langle z \rangle = 1.0$, the dipole is much less prominant and the $\kappa$ and $\kappa_\delta$ signals are more similar by eye.
Qualitatively, this aligns with our expectations that peculiar velocity effects (captured by the Doppler lensing) contributing to anisotropies in distances are more prominent at lower redshifts and decay at higher redshift. 

\begin{figure}%
    \centering
    \includegraphics[width=\columnwidth]{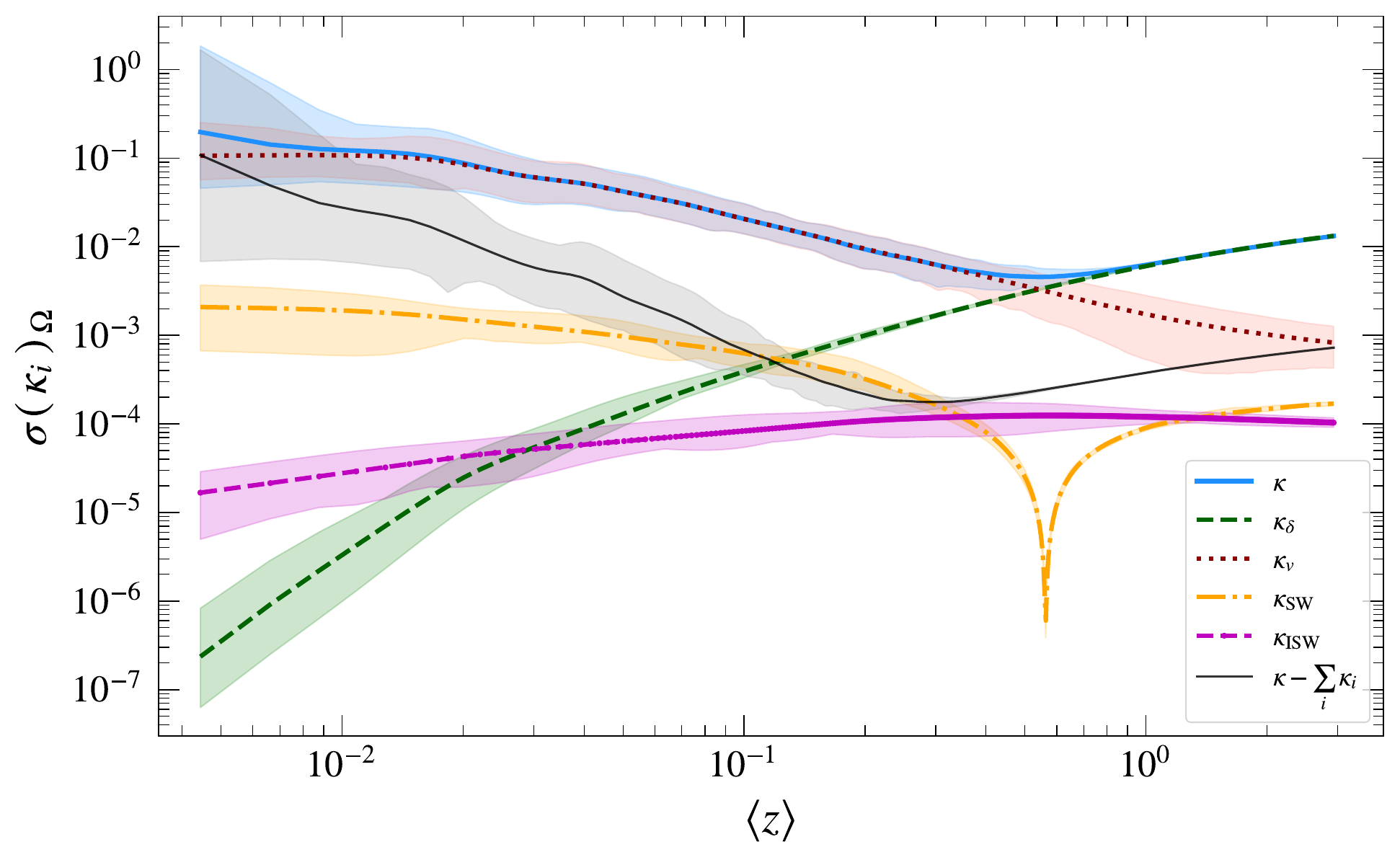}
    \caption{Standard deviation of fluctuations in the lensing convergence across the sky as a function of mean redshift. We show the nonlinear signal, $\kappa$ (blue solid curve), as well as the linearised contributions $\kappa_\delta$ (green dashed curve), $\kappa_v$ (red dotted curve), $\kappa_{\rm SW}$ (orange dot-dashed curve), and $\kappa_{\rm ISW}$ (magenta dashed curve). The difference between the nonlinear convergence and all of the combined linearised contributions is shown by a thin black solid curve. Each line represents the standard deviation across the sky averaged over 20 observers in the $N=256$ simulation. Shaded regions around each curve show the full range of variance across all observers.}
    \label{fig:kappastds}
\end{figure}

Notably, the chosen slice of $\langle z \rangle = 0.5$ in the top panels of Figure ~\ref{fig:skymaps} corresponds to the redshift where the contributions of $\kappa_\delta$ and $\kappa_v$ are roughly similar in magnitude. This is more easily visible in Figure~\ref{fig:kappastds}, which shows the standard deviation of the fluctuations in each convergence signal across the sky as a function of redshift.
We show the nonlinear convergence, $\kappa$, with a solid blue curve, and all linearised contributions are shown with broken curves. The linearised convergence from the density, $\kappa_\delta$, is the green dashed curve, the Doppler lensing $\kappa_v$ is the red dotted curve, and the SW and ISW contributions are the yellow dot-dashed and magenta dashed curves, respectively. Curves show the mean over all 20 observers and shaded regions indicate the full spread across the observers. The thin black solid curve shows the standard deviation of the difference map $\kappa - \sum_a \kappa_a$ where $\kappa_a=[\kappa_\delta, \kappa_v,\kappa_{\rm SW},\kappa_{\rm ISW}]$. %

Figure~\ref{fig:kappastds} gives us a rough idea of the relative size of each contribution to the fully nonlinear signal $\kappa$. As expected, the Doppler contribution (red dotted curve) is dominant at low redshifts, only to be surpassed by the linearised density convergence, $\kappa_\delta$, at $z\approx$ 0.6. This aligns with expectations from earlier works \citep[e.g.][]{Bacon:2014}. Also as expected, the SW and ISW contributions are subdominant at large redshifts. However the SW (ISW) fluctuations are larger in amplitude than $\kappa_\delta$ for $z\lesssim$ 0.1 (0.03). Note the dip in the SW variance occurs at the redshift where the two terms proportional to $\phi_s$ in \eqref{eq:kappasw} are approximately equal (i.e., for EdS $(1+z_s)/H\chi_s \approx 2$ at $z\approx 0.56$).
One may be surprised to see the amplitude of the ISW contribution of $\mathcal{O}(10^{-4}$--$10^{-5})$ in our matter-dominated simulations. We remind the reader that the ISW contribution also contains an integral of $\phi$ and is not purely sourced by $\phi '$. In fact, $\phi '$ is at least three orders of magnitude smaller than $\phi$ for all redshifts we study (see Appendix~\ref{appx:rel_calcs}). 

All known relativistic corrections combined (black solid curve in Figure~\ref{fig:kappastds}) approximate the amplitude of fluctuations in the nonlinear convergence to better than $\sim 1$\% for $z\gtrsim 0.02$. For redshifts $z<0.01$, the nonlinear convergence signal is a few to 10\% different (and up to 100\% for some observers) than the sum of the linear contributions. 
However, for lower redshifts of $z\lesssim 0.2$, the fluctuations for all contributions on a given light-cone slice are not well represented by a Gaussian, and so the standard deviation is not necessarily a reliable measure of typical fluctuations at these small scales. 
In the next section, we will make a more robust comparison of the linearised and nonlinear signals via the use of angular power spectra at a series of redshifts.

\subsection{Angular power spectra}\label{sec:results_Cls}

Assessing the angular power spectra of both the nonlinear and linearised convergence signals allows us to precisely quantify how well the approximations fair at a variety of redshifts and angular scales. We will calculate the auto-correlation function $C_{\kappa\kappa} (\ell)$, where $\ell$ is the angular multipole, using the \texttt{healpy.anafast} Python function \citep{Zonca:2019}. Note that we sometimes refer to the auto correlation function as, e.g., $C_{\kappa}$ for brevity.

An important limitation on observations of power spectra, especially on the large angular scales we study, is the
so-called \textit{cosmic variance} limit. 
This is an observational uncertainty which arises from the fact that our Universe is only a single realisation of a particular theoretical power spectrum---it contains only a limited number of large-scale spatial fluctuations (i.e., in $k$-space). A different set of large-scale $k$ modes may well induce a larger change in the observed angular power spectrum than a change in the underlying theoretical model itself. For all-sky data observed on a single light-cone slice, this amounts to an uncertainty of $\Delta C_\kappa / C_\kappa = \sqrt{2/(2\ell+1)}$. For such a scenario, as we consider in this work, this error is irreducible regardless of the precision of the observations themselves. Though, some combinations of observed spectra may be able to reduce this, see e.g. \citet{Seljak:2009}. Additionally, tomographic binning of the data in redshift \citep{Wu:1999} can improve the cosmic variance limitation for weak lensing analyses \citep{Pen:2004}.

\begin{figure}
    \centering
    \begin{minipage}[t]{\linewidth}
        \centering
        \includegraphics[width=\linewidth]{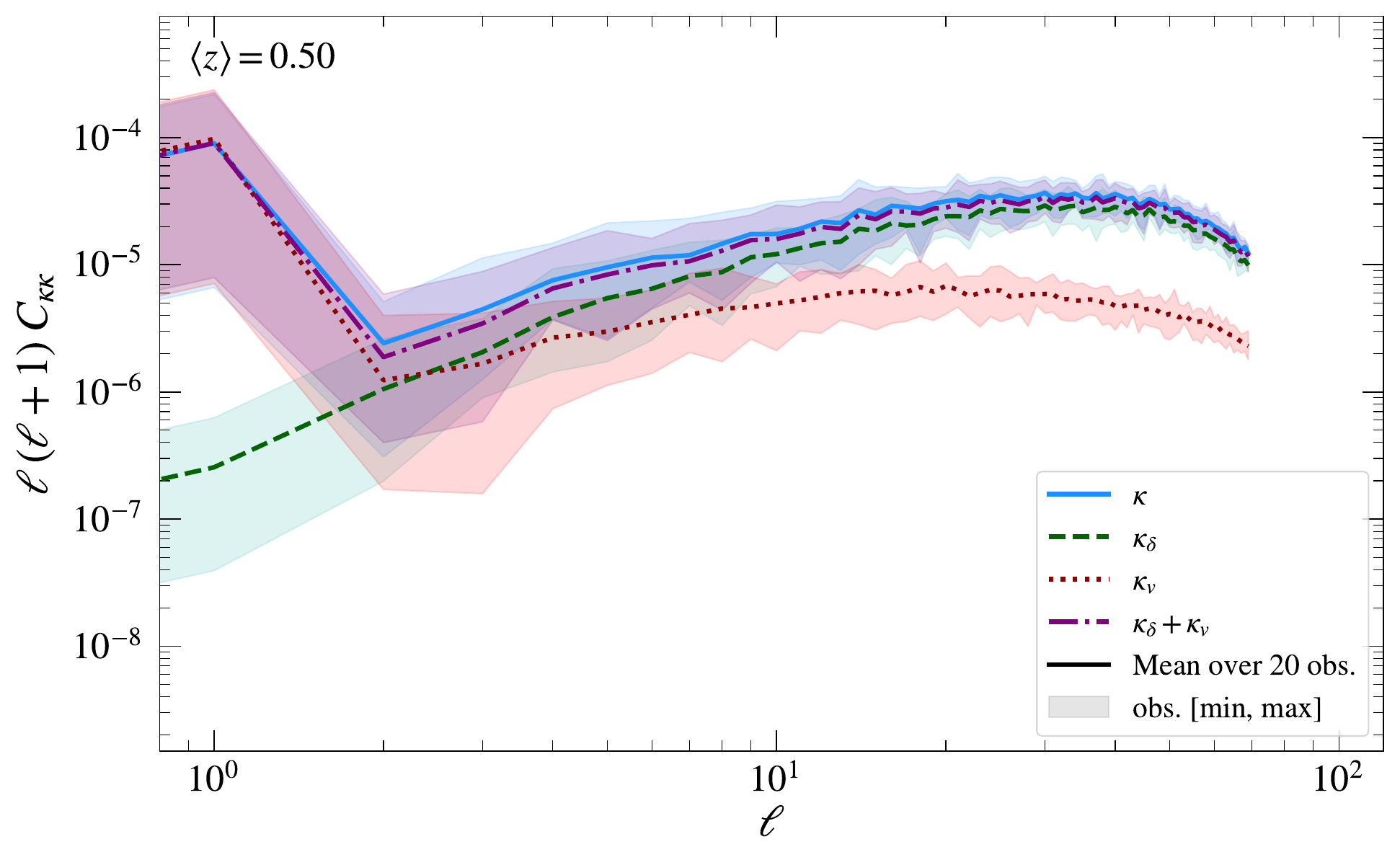}
    \end{minipage}
    \begin{minipage}[t]{\linewidth}
        \centering
        \includegraphics[width=\linewidth]{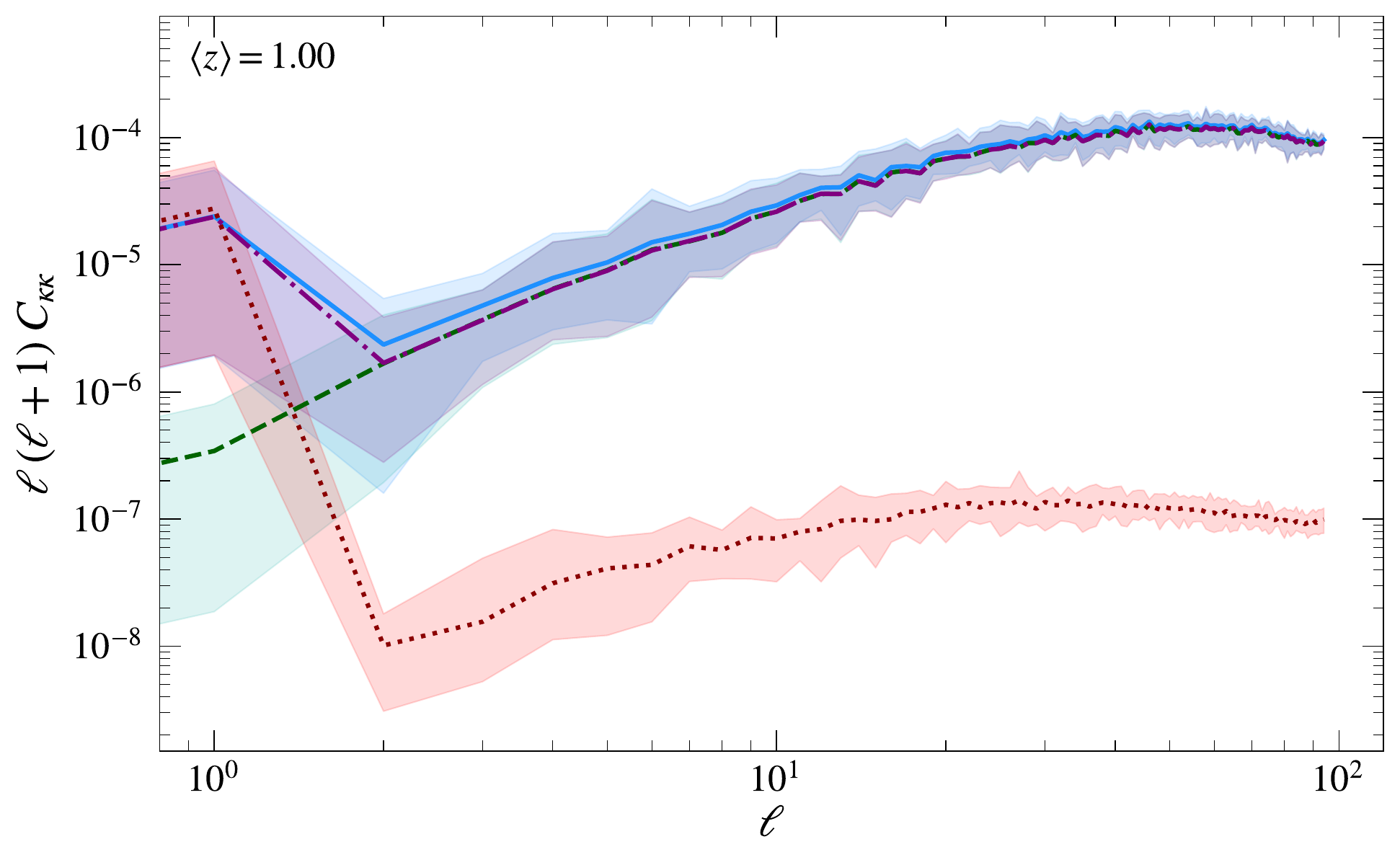} 
    \end{minipage}
    \caption{Angular power spectra (averaged over 20 observers) of the relativistic convergence $\kappa$ (solid curves), the density convergence $\kappa_\delta$ (dashed curves), and the Doppler lensing $\kappa_v$ (dotted curves) as a function of angular scale, $\ell$, for two slices with mean redshift $\langle z \rangle = 0.5$ (top) and 1.0 (bottom). The shaded regions show the range within which all 20 observers' power spectra fall.}
    \label{fig:Cls_comp}
\end{figure}

\begin{figure*}
    \centering
    \begin{minipage}[t]{0.49\textwidth}
        \centering
        \includegraphics[width=\linewidth]{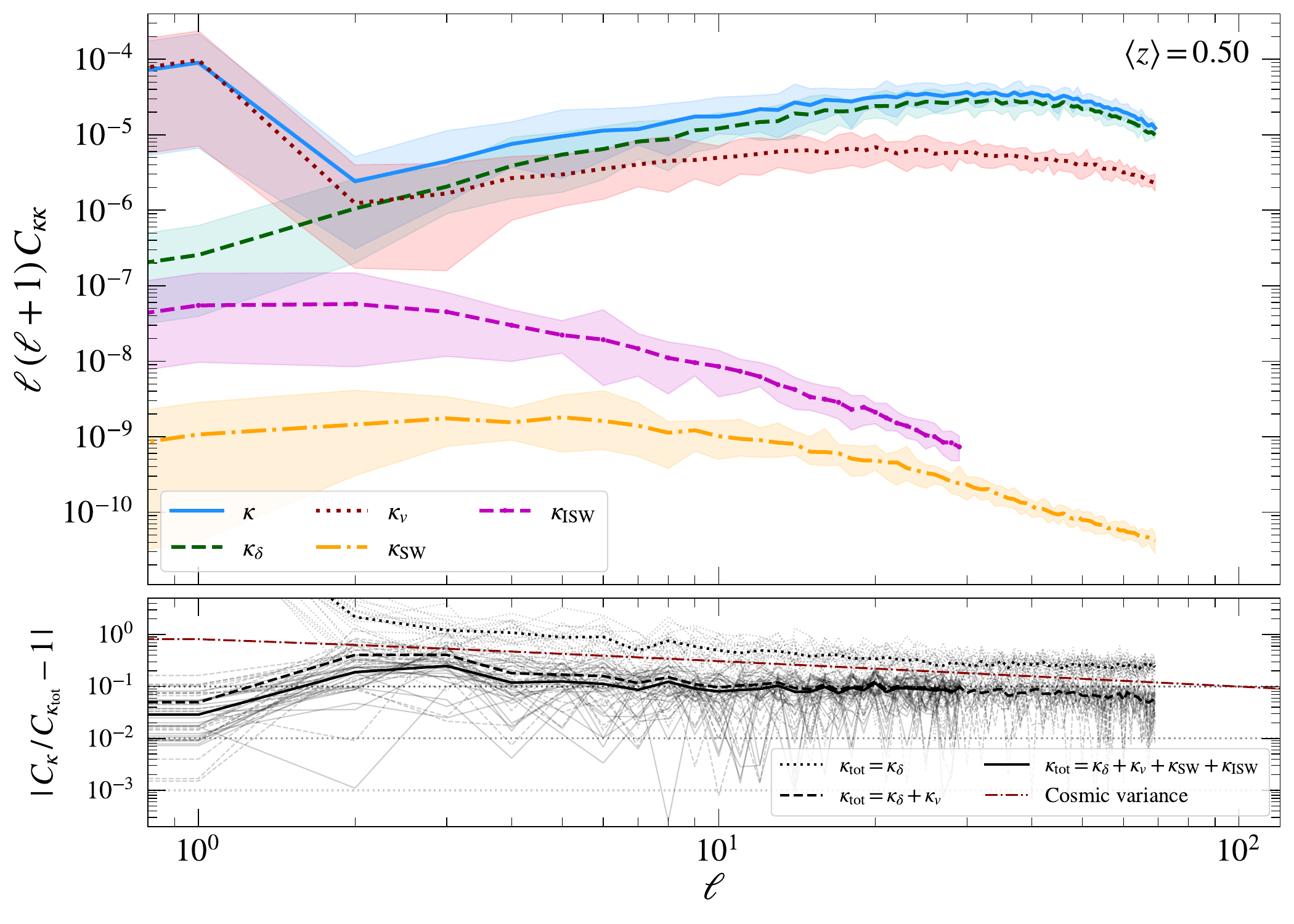}
    \end{minipage}
    \begin{minipage}[t]{0.49\textwidth}
        \centering
        \includegraphics[width=\linewidth]{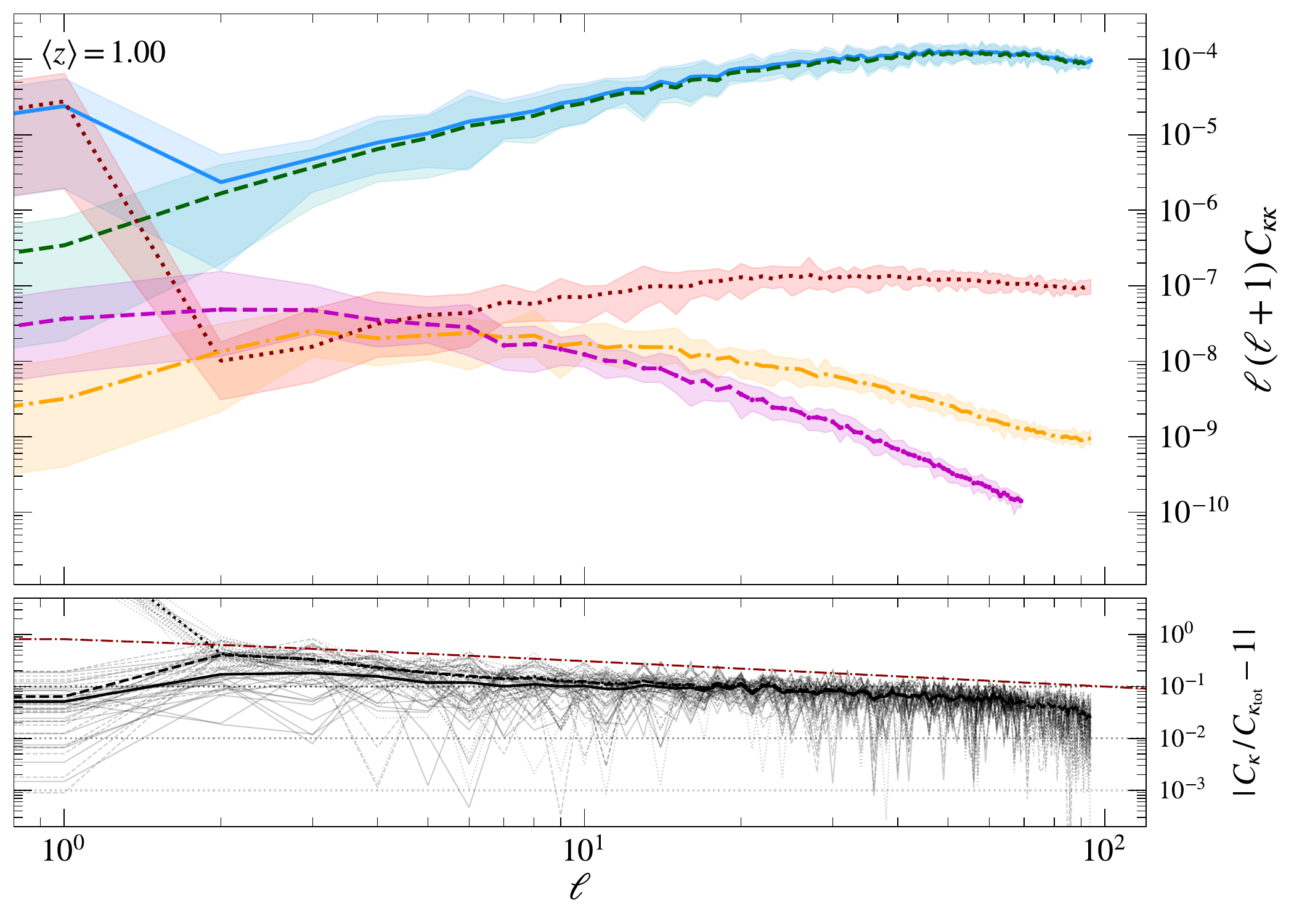} 
    \end{minipage}
    \caption{Angular power spectra of the nonlinear convergence (blue solid curves), linearised convergence from the density (green dashed curves) as well as all known relativistic contributions. The Doppler lensing is shown by the red dashed curves, the SW is the orange dot-dashed curves, and the ISW contribution is the magenta dashed curves. Top panels show the power spectrum averaged over all 20 observers with shaded regions showing the range of minimum to maximum over all observers. Bottom panels show the absolute difference in the power spectra when comparing $\kappa$ and $\kappa_\delta$ (dotted curves), $\kappa$ and $\kappa_\delta + \kappa_v$ (dashed curves), and $\kappa$ and $\kappa_\delta + \kappa_v + \kappa_{\rm SW} + \kappa_{\rm ISW}$ (solid curves). In the bottom panel, thick curves are the mean over 20 observers and thinner curves show individual observers, with red dot-dashed curves showing the cosmic variance limit as a function of $\ell$. Left panels show slices with $\langle z \rangle = 0.5$ and right panels show $\langle z \rangle = 1.0$.}
    \label{fig:Cl_comp_all}
\end{figure*}
Figure~\ref{fig:Cls_comp} shows the angular power spectra for the nonlinear convergence (blue solid curve), the linearised convergence from the density field (green dashed curve), and the Doppler lensing (red dotted curve). The purple dot-dashed curve shows the power spectrum of the total signal $\kappa_\delta + \kappa_v$.
Top panel shows the redshift slice $\langle z \rangle = 0.5$ and the bottom panel shows $\langle z \rangle = 1.0$. Curves indicate the mean power spectra over all 20 observers\footnote{Our `average' power spectra are always $C_\kappa$ averaged over all observers and \textit{not} the power spectra of the $\kappa$ sky-map averaged over observers.} and shaded regions represent the full range of variation across all observers. 
The top panel is truncated at $\ell>70$ since we find at lower redshift the power spectra are not numerically converged on these smaller angular scales (see Appendix~\ref{appx:convergence}).

From Figure~\ref{fig:Cls_comp}, we can see at lower redshifts and larger angular scales that $\kappa_\delta$ fails to solely reproduce the nonlinear signal $\kappa$ \citep[as expected from previous works; e.g.][]{Bonvin:2006,Bacon:2014,Fleury:2017}. At $\langle z\rangle = 1$, 
the Doppler lensing is negligible for all $\ell > 1$. 
Next we will consider whether including the SW and ISW effects can further improve the approximation to $\kappa$.

Figure~\ref{fig:Cl_comp_all} shows the angular power spectra of all convergence signals we calculate in this work at two redshift slices $\langle z \rangle=0.5$ (left panels) and $\langle z \rangle = 1.0$ (right panels). Solid blue curves show the nonlinear convergence $\kappa$, green dashed curves show the linearised convergence from the density $\kappa_\delta$, red dotted curves show the Doppler lensing $\kappa_v$, orange dot-dashed curves show the SW lensing, and magenta dashed curves with points show the ISW lensing. Curves in the top panels are angular power spectra averaged over 20 observers and associated shaded regions indicate the minimum to maximum range of power across all observers. Bottom panels show the absolute value of the normalised difference between the nonlinear signal and various sums of the linearised contributions (as indicated in the legend), as well as the cosmic variance limit (red dot-dashed curve). 
Note that for each power spectrum we only show angular scales that are numerically converged, which are the same for all contributions except for the ISW signal (see Appendix~\ref{appx:convergence}).

The main conclusion we draw from Figure~\ref{fig:Cl_comp_all} is that the SW and ISW contributions are far subdominant to the other contributions, as expected based on perturbation theory predictions. To quantify this, including the SW and ISW terms in approximating $\kappa$ makes about a factor of $\sim 2$ improvement for large scales $\ell\lesssim10$, and no difference for smaller scales (when considering the mean across observers; moving from dashed to solid curves in the lower panels). However, some observers enjoy up to a $10\times$ improvement at large scales $\ell<10$, while for some observers the fit becomes worse with the inclusion of these terms. 

For almost all cases (with the exception of a few observers for $\ell=2,3$ at low redshift; see also Section~\ref{sec:Clzevol} below)
the difference between the linearised and nonlinear convergence---when including all known GR effects---is below the cosmic variance limit (red dot-dashed curve in the bottom panels). When only considering the convergence from the density field (dotted curves in the bottom panels) at lower redshift (left panel), the difference is above cosmic variance for most observers for all $\ell$. For higher redshift (right panel), it drops below this level for all $\ell>1$.

\subsection{Redshift evolution of the difference}\label{sec:Clzevol}

In the previous section we have focused on two particular redshift slices of $\langle z \rangle = 0.5$ and 1.0. In this section, we will investigate how the difference in the power spectra evolves as a function of redshift for different angular scales. 

\begin{figure*}
    \centering
    \begin{minipage}[t]{0.49\textwidth}
        \centering
        \includegraphics[width=\linewidth]{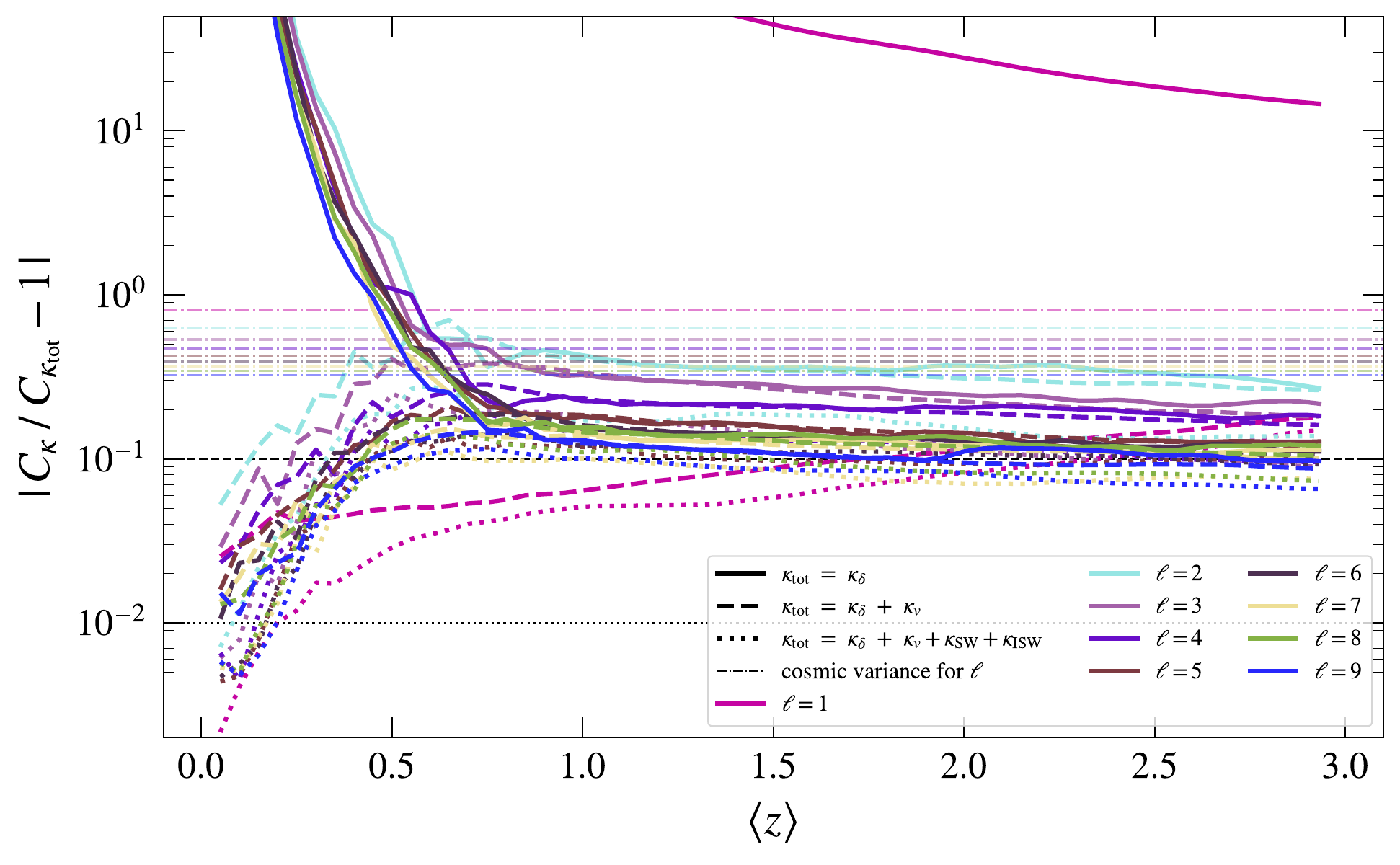}        
    \end{minipage}
    \begin{minipage}[t]{0.49\textwidth}
        \centering
        \includegraphics[width=\linewidth]{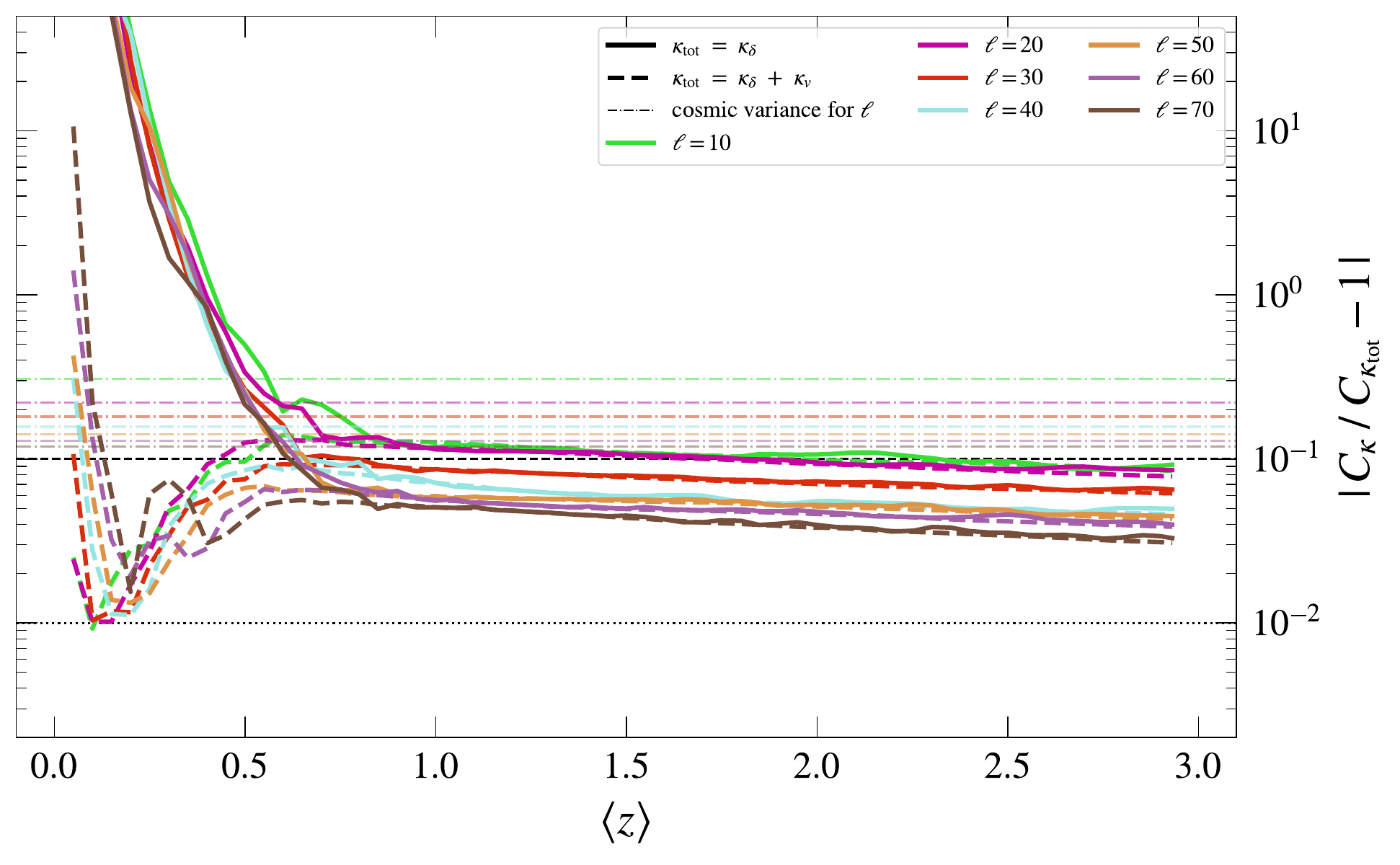}
    \end{minipage}
    \caption{Redshift evolution of the relative difference in power spectra between nonlinear and linearised convergence. Solid curves show the difference when only including $\kappa_\delta$, dashed curves show the difference when including $\kappa_\delta + \kappa_v$, and dotted curves (left panel only) show the difference when including the SW and ISW contributions. Dot-dashed horizontal lines show the cosmic variance limit for the $\ell$ of the same colour. Left panel shows $\ell<10$ and right panel shows $\ell\geq10$. }
    \label{fig:Cldiff_vs_z}
\end{figure*}
Figure~\ref{fig:Cldiff_vs_z} shows the relative difference in power spectra as a function of redshift for large angular scales (left panel) and smaller angular scales (right panel), averaged over 20 observers.
More specifically, we show the difference between the nonlinear convergence power spectrum, $C_\kappa$, and the power spectrum of the approximate map when including different relativistic contributions, $C_{\kappa_{\rm tot}}$. Solid curves are when we only include the linearised convergence from the density field, $\kappa_{\rm tot}=\kappa_\delta$, dashed curves when we also include the Doppler lensing, $\kappa_{\rm tot}=\kappa_\delta+\kappa_v$, and dotted curves when further including SW and ISW. Different coloured curves are different values of $\ell$ as indicated in each legend. 
Black horizontal lines in each panel indicates a 10\% and 1\% difference for reference, and coloured horizontal dot-dashed lines show the cosmic variance limit for the $\ell$ of the same colour. In the left panel, the amplitude of the dipole difference for $\kappa_{\rm tot}=\kappa_\delta$ (solid magenta curve) reaches $\sim 10^6$ at low redshift and the other curves reach $\sim 10^5$. 
As in the previous section, we only present curves for angular scales and redshifts that are numerically converged (see Appendix~\ref{appx:convergence}).

In Figure~\ref{fig:Cldiff_vs_z} we see the drastic improvement when including peculiar velocities at low redshift for all scales (moving from solid to dashed curves in both panels). However, we still see a $\sim$10\% difference in the power spectra for low-$\ell$ even when including all known relativistic corrections (dotted curves in the left panel). 
For $\langle z \rangle > 0.8$, the inclusion of peculiar velocities to the lensing makes no significant improvement to the fit to the nonlinear convergence for all angular scales $\ell \leq 70$. 
For $1< \langle z \rangle < 3$, overall there is a persistent few-to-10\% difference in power spectra; which reaches 20--30\% for the largest scales. 

As in the previous section, we also find that for almost all $\ell$ and $\langle z \rangle$ the differences we find are below cosmic variance (horizontal coloured lines in Figure~\ref{fig:Cldiff_vs_z}). 
We remind the reader that these curves are the \textit{average} over 20 independent observers. If we instead consider the \textit{maximum} over our observers we find that the quadrupole difference ($\ell=2$; cyan curves) surpasses the cosmic variance for $z\approx$~0.3--1.5 even when including the Doppler contribution. This also applies to the octupole ($\ell=3$) for $z>0.4$. 
For both of these cases, including the SW and ISW contributions reduces the difference below cosmic variance except for $z\approx 0.5$. 
For higher $\ell$ (right panel) and all $z\gtrsim 0.2$ in all cases the difference (when including the Doppler term) is below cosmic variance also for the maximum across observers. For the density contribution only, this is the case for all $z>1$. 
Our sample of 20 observers is relatively small, and if we were to ray trace for a larger sample of observers we may well see more cases for which the difference is above the level of cosmic variance. We also remind the reader that the cosmic variance shown in Figure~\ref{fig:Cldiff_vs_z} and Figure~\ref{fig:Cl_comp_all} is specific to a light-cone slice and may be reduced by, e.g., tomographic binning of our mock data. We also anticipate the amplitude of the difference may change when considering redshift bins rather than slices. 
We leave such an investigation to future work.

\subsection{Individual lines of sight}\label{sec:LOS}

So far we have considered statistics of the lensing convergence signal on the observers' simulated skies. This is useful for cosmological surveys which sample an area on the sky and can thus measure the angular power spectrum, i.e. weak lensing surveys or CMB measurements. 
However, we might also be interested in how well the linearised convergence fares along individual lines of sight. 

This is relevant for observations which consider individual objects rather than a particular portion of the sky; for example high-redshift supernovae and strong lensing. In these cases, a number of point sources are considered to constrain cosmological parameters via their inferred magnitudes and redshifts \citep[in the case of supernovae; e.g.][]{Popovic:2025} or delays in image arrival time \citep[in the case of strong lensing; e.g][]{Birrer:2020}. All geodesics traversing an inhomogeneous space-time will feel the effects of lensing due to intervening structures. For point sources, this can be treated as an additional systematic via a predicted scatter in the Hubble diagram \citep{Holz:2005}. 
Alternatively, the lensing of supernovae can be detected and used to constrain cosmological parameters and---in the future---to study the distribution of mass in the Universe \citep{Shah:2024,Macaulay:2020,Holz:1998}.
Such approaches adopt the common approximation \eqref{eq:kappaapproxdef} in theoretical modeling of the lensing. %

Strong lensing time-delay measurements typically adopt a single-plane approximation to describe the system of observed multiple images \citep[see, e.g.][for a review]{Birrer:2024,Treu:2022}; which is distinct from the framework adopted in our ray tracing. 
However, the images of a strongly-lensed system are still magnified or demagnified due to intervening structure along the line of sight as an {additional effect}. This can be included in the theoretical modeling of the time delay and constrained along individual lines of sight \citep[with the use of \eqref{eq:kappaapproxdef} and informed by Newtonian simulations; see][]{Birrer:2020}.

\begin{figure}
    \centering
    \includegraphics[width=\linewidth]{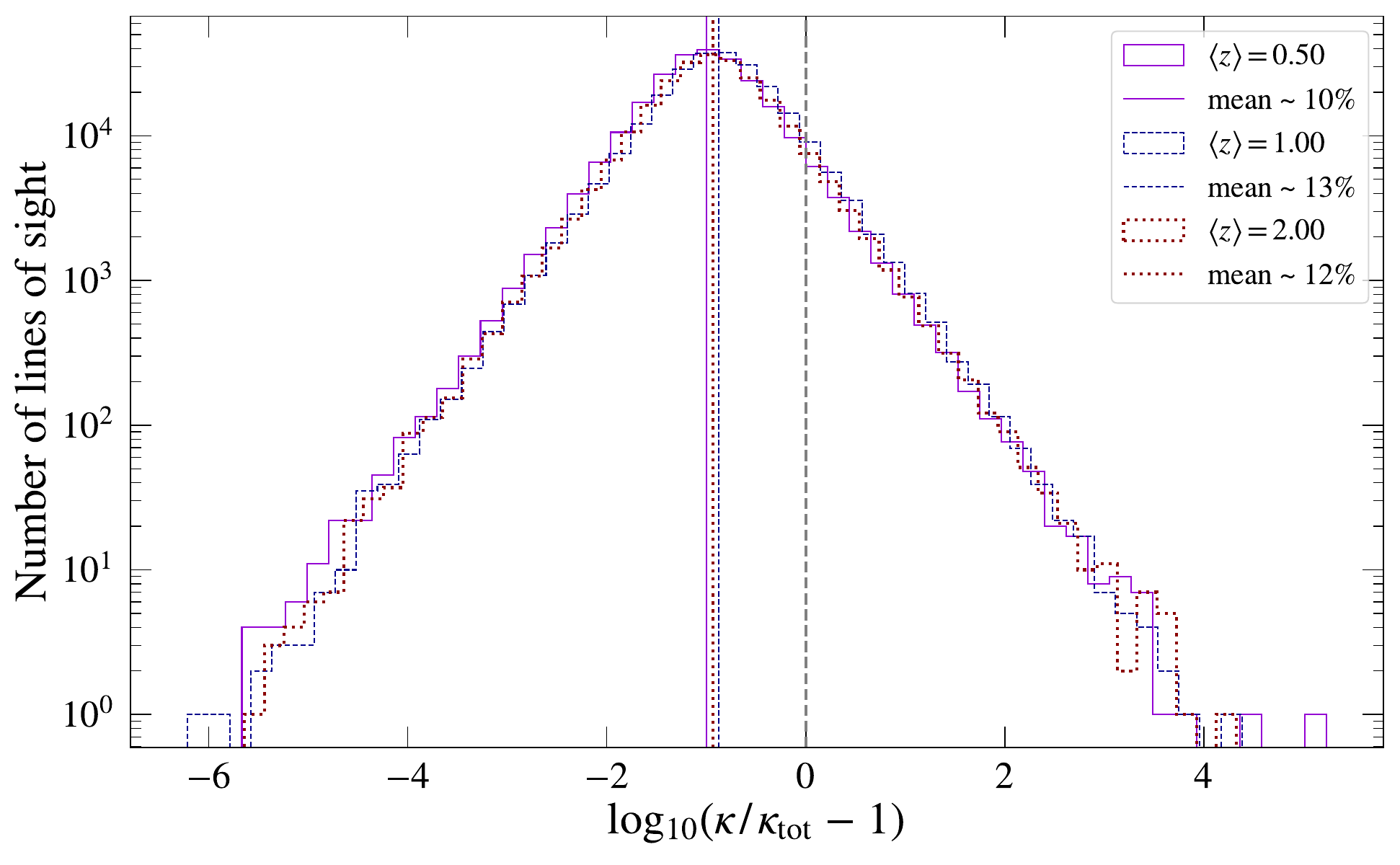}
    \caption{Histogram of the relative difference between the nonlinear convergence, $\kappa$, and the linearised convergence, $\kappa_{\rm tot}=\kappa_\delta+\kappa_v + \kappa_{\rm SW} + \kappa_{\rm ISW}$. We include all lines of sight for all 20 observers in the same simulation for redshift slices $\langle z \rangle = 0.5$ (solid pink), $\langle z \rangle = 1$ (dashed blue), and $\langle z \rangle = 2$ (dotted red). Mean values are shown as vertical lines.}
    \label{fig:kdiff_all_los}
\end{figure}
Figure~\ref{fig:kdiff_all_los} shows the relative difference between the nonlinear and linearised convergence for all lines of sight for all 20 observers. Different histograms are for different redshift slices: $\langle z \rangle = 0.5$ (solid pink), $\langle z \rangle = 1$ (dashed blue), and $\langle z \rangle = 2$ (dotted red). Vertical lines are the mean of the histogram of the same style and colour, showing the $\mathcal{O}(10\%)$ difference on average for each redshift slice. We find that for 6\% (8\%) of all lines of sight, the linearised convergence fails by more than 100\% at $\langle z \rangle=0.5$ (1 and 2).

While we cannot identify individual lines of sight in the simulation with those which may be coincident with strong lensing events or supernovae, the potential for very large differences shown in Figure~\ref{fig:kdiff_all_los} motivates further study. In particular for the potential bias on cosmological parameters as measured from a small number of (or individual) objects---for example; via strongly-lensed supernovae \citep[e.g.][]{Pierel:2026} or ``bright siren'' gravitational wave events \citep{Mpetha:2024}.

\subsection{Comparison to existing works}\label{sec:comparison}

Several previous works have studied relativistic effects in the lensing power spectrum. In the introduction, we mentioned \citet{Giblin:2017a} as the first quantification from an NR simulation. In their Figure~3, the authors find a $\sim$2--3\% difference between the relativistic and linear convergence power spectra for $z=0.25$ and $\ell=$~10--30. Comparing with these same multipoles in the right panel of our Figure~\ref{fig:Cldiff_vs_z} (dashed curves, since \citet{Giblin:2017a} also include the Doppler term) we find a difference of 6--10\% (variance across $\ell=10,20,30$), a factor of a few larger than their results.
For larger angular scales of $\ell=$~2--10, \citet{Giblin:2017a} find differences of 3--30\% that decrease with increasing $\ell$. Comparing with the dashed curves in the left panel of our Figure~\ref{fig:Cldiff_vs_z}, we instead find a $\sim$10--60\% difference at this same redshift, which also decreases with increasing $\ell$. This discrepancy in results may be due to different methods in calculating the `background' quantities. For our fiducial analysis, we use the background scale factor and co-moving distance calculated using the observed redshift. If we instead use these quantities as calculated on the simulation hypersurface, our results are more consistent with \citet{Giblin:2017a} (see Appendix~\ref{appx:chi_Born}).

\citet{Lepori:2020} studied relativistic corrections to the lensing convergence power spectrum using the \textit{gevolution} $N$-body code in the weak-field limit. Since this work used large-scale particle simulations and a pencil-beam ray tracing (rather than full-sky as we adopt here), they were able to study much smaller angular scales than we did in this work. However, the authors maximum angular scale of $\ell=50$ does overlap with our work. For $z=1.5$ (as shown in their Figure~10), the authors find that linear theory is accurate to $<1\%$ (and a difference that is consistent with zero within their error bars). We can compare this with the orange curve in the right-panel of Figure~\ref{fig:Cldiff_vs_z}, representing the relative difference in power spectra with respect to the linear approximation for $\ell=50$ as a function of redshift. At $z=1.5$, we find a difference of $\sim 5\%$, slightly larger than \citet{Lepori:2020} yet still within their error bars. %

We note that the work of \citet{Lepori:2020} used simulations that are based explicitly on a metric of similar form as \eqref{eq:pertmetric} (though with vector and tensor modes included), and so their work will not suffer from potential issues in identifying perturbations themselves (since these are clearly defined and explicitly evolved in their simulations). \citet{Giblin:2017a} also adopt NR simulations and so their work is more directly comparable to ours in this sense. 

\citet{Rasera:2022} calculate a number of relativistic effects in cosmic observables via ray tracing in Newtonian $N$-body simulations. While the ray tracing adopts several relativistic effects (defined as anything except the standard $\kappa_\delta$ as defined in \eqref{eq:kappaapproxdef}), the simulated fields are still Newtonian and the calculations adopt the perturbative form of the metric tensor as in \eqref{eq:pertmetric}.
The authors find a significant 20\% (30\%) difference for $z=1.8$ ($z=0.7$) between the relativistic-corrected Newtonian results and the CLASS\footnote{http://class-code.net} (linear) power spectrum at large angular scales of $\ell\sim$~30--40. This difference is larger than what we find for these redshifts and angular scales: in the right panel of Figure~\ref{fig:Cldiff_vs_z} we find $\sim 10$\% ($\sim 5\%$) for $z\approx0.7$ ($z\approx1.8$) for $\ell=30$ and 40. The authors of \citet{Rasera:2022} attribute the difference they find to the finite size of their simulated light cone, which could also be the source of the difference between their result and ours. There are of course several key differences in simulation and analysis between our work and \citet{Rasera:2022} that make a direct comparison difficult. 

Overall, our results fit reasonably into the scope of these existing studies. Importantly, we expand the large-angle multipole study of \citet{Giblin:2017a} to more redshifts, angular scales, and observers, and confirm the large-angle result of \citet{Lepori:2020} in a NR framework.

\subsection{Reasons for the remaining discrepancy}\label{ssec:why}

The linearised contributions in general provide a very good fit to the nonlinear convergence; especially when including the effect of peculiar velocities at low redshifts. However, a difference in power spectra of a few to 10\% ($\ell$-dependent) persists---even when including all known general relativistic corrections---for the full redshift range of $0.05<z<3$ that we study. %
Importantly, this difference is not observable due to cosmic variance (for a redshift slice). However, we are still interested to investigate possible causes of the discrepancy for the particular realisation of our simulated universe. 

All results we have presented in the main text are robust to changes in numerical resolution, which we show in Appendix~\ref{appx:convergence}, and are not dominated by constraint violation, which we show in Appendix~\ref{appx:constraints}. %

As briefly discussed in Section~\ref{sec:kappa_approx}, in deriving $\kappa_\delta$ from $\kappa$, several terms of linear order in derivatives of $\phi$ have been neglected, and the Poisson equation, $\nabla^2\phi \propto \delta$, has been assumed valid. In Appendix~\ref{appx:kappa_terms}, we calculate these neglected terms and show that their inclusion makes no noticeable difference to our results. We also show that the Poisson equation is mostly sufficient in our simulations---with an interesting deviation at lower redshift and smaller angular scales that perhaps requires further investigation---and makes no significant change to our main results.

\citet{Umeh:2014} studied corrections to the lensing convergence at second order in perturbation theory, which includes some cross-terms between the first-order $\kappa_\delta$ and $\kappa_v$ which can be significant. However, based on the estimates given by the authors that these corrections are relevant only for very large density contrasts (see their eq~(48)), we expect them not to contribute significantly in our simulations.

A key assumption in our calculation of all the linearised convergence signals is that the metric tensor from our NR simulation retains its initial decomposition in the form of \eqref{eq:pertmetric}; with the same EdS background as we specified for the initial data. Due to the development of nonlinearities and the fact that NR does not enforce an FLRW background evolution, at late times the space-time metric is not guaranteed to follow the EdS solution (or, in fact, any FLRW model) in terms of averages \textit{or} observationally inferred parameters. Even a slight deviation from an `average' EdS-like evolution in our simulation could spill into the extraction of perturbations and thus yield a difference between our nonlinear $\kappa$ and the linearised contributions. 
Our motivation for assuming the EdS model aligns with what we typically assume in standard cosmology. Specifically, that fluctuations in $\gamma_{ij}$ remain `small' at late times in the simulation (see Appendix~\ref{appx:kappa_terms}) and thus the best-fit background cosmological model does not change significantly\footnote{This condition alone is not sufficient to guarantee the model universe will be close to any FLRW model; see \citet{Rasanen:2010}, though this assumption is common throughout standard cosmology}.
Our assumption is further motivated by the fact that large-scale, spatially-averaged qualities of the simulation have been shown to remain close to EdS even at late times \citep[see][for analysis of similar simulations]{Macpherson:2019}.

However, some works have found inferred FLRW parameters including the Hubble constant \citep{Macpherson:2024} and density parameters \citep[][in prep.]{Elgamal_inprep} can be up to a few percent different from the EdS model. We thus might consider if a different set of FLRW parameters could improve the fit to the nonlinear convergence. 
Due to the observer-dependence of the best-fit cosmological parameters found in the aforementioned works, it is beyond the scope of this paper to explore all possible changes in the background FLRW model and the consequent effect on our results. Defining a `background' for a simulation that is non-perturbative (or, indeed, the Universe itself) is non trivial and related to the unsolved \textit{fitting problem} in cosmology \citep{Clarkson:2011}.

\subsection{Caveats}\label{sec:caveats}

Some important caveats to our results are briefly discussed here. 

We consider several numerical aspects of the simulations to be limiting our ability to make conclusive statements on these effects. 
Firstly, we adopt a continuous fluid approximation for the dark matter in our model universe. Dark matter is best approximated as a set of collisionless particles which allow for the formation of bound structures (i.e. galaxies and galaxy clusters). Several works have incorporated $N$-body particles alongside NR \citep{Giblin:2019,East:2019,Daverio:2019}, however, the incorporation in the ET \citep{Magnall:2023} is not yet ready for large-scale realistic simulations. 
An important aspect of the continuous fluid approximation is that it limits the minimum physical scale we can reliably simulate in our simulations. We strictly limit the minimum scale to $\sim 24\, h^{-1}$ Mpc via the physical size of our grid cells. The result is our simulations are missing \textit{all} structures below this scale. Additionally, power in structures below the scale of $\sim 10$ grid cells is \textit{damped} in order to reduce numerical noise; which further reduces the inhomogeneity. A higher degree of nonlinearity in the simulations would likely impact the results. 

Secondly, our simulations do not contain a cosmological constant. In generating our initial data, we have used the \lcdm\, matter power spectrum at $z=0$ such that the local environments of our observers are as similar as possible to observations. However, the simulation itself is evolved without $\Lambda$ and thus our results will be different to those obtained in a model universe with $\Lambda$.

Additionally to these numerical aspects, as we have discussed in Section~\ref{ssec:why}, %
we assume that the initial EdS model remains a good fit via the adoption of the metric \eqref{eq:pertmetric}. This is required to extract the linearised convergence signals from the simulation. 
While this is accurate in the sense of large-scale spatial averages \citep{Macpherson:2019}, observationally the best-fit FLRW parameters can be observer dependent and different from EdS \citep{Macpherson:2021,Elgamal_inprep}. 

Due to the high computational expense of the ray tracing analysis we are limited in our current study to only 20 observers in one simulation. Our results thus would likely change if we considered a larger number of observers as well as simulations with different realisations of the initial power spectrum.

\section{Conclusions}\label{sec:conclude}

In this work we have studied the convergence signal calculated in an end-to-end nonlinear general relativistic framework of numerical relativity simulations and ray tracing. For a set of 20 observers, we compared this nonlinear convergence to the expectation based on perturbation theory. Our main conclusions from the study of angular power spectra, as averaged over 20 observers, can be summarised as follows:
\begin{itemize}
    \item The Doppler signal makes a significant contribution to the  convergence for redshifts $z\lesssim0.6$ for all angular scales we study.
    \item The linearised convergence proportional to $\delta$ reproduces the nonlinear convergence within 10\% at $z\gtrsim$0.6 for $\ell=$~30--70 and $z\gtrsim 1.5$ for $\ell=$~10--20. For $\ell<10$ the difference is 10--30\% for all redshifts.
    \item When considering all known general-relativistic corrections, we find a residual 3--30\% difference between the linearised and nonlinear convergence signals at all redshifts (dependent on angular scale).
\end{itemize}

All of these conclusions are robust to changes in numerical resolution of the simulation and ray tracing. 
In almost all cases, the differences we find are below the level of cosmic variance for observations on a redshift slice. While we cannot  attribute the difference to any particular source, we suggest the most likely reasons to be either 1) the difficulty
in assigning a perturbative description to our non-perturbative simulation or 2) the limited number of observers and/or simulations in our sample. 

Future work could include investigating whether the differences we find are above the cosmic variance limit for a tomographic lensing survey, as well as pushing to smaller angular scales. The latter would most likely require using cosmological numerical relativity simulations which adopt $N$-body particle dynamics so that the simulations themselves also accurately represent smaller physical scales.

\begin{acknowledgments}

HJM would like to thank Yuuki Omori, Daniel Gilman, and the Survey Science LSS subgroup at UChicago for helpful discussions related to this work.
Support for HJM was provided by NASA through the NASA Hubble Fellowship grant HST-HF2-51514.001-A awarded by the Space Telescope Science Institute, which is operated by the Association of Universities for Research in Astronomy, Inc., for NASA, under contract NAS5-26555. HJM was also supported by the Kavli Institute for Cosmological Physics through an endowment from the Kavli foundation and its founder Fred Kavli. Some of the results in this paper have been derived using the \texttt{HEALPix} \citep{Gorski:2005} package. Simulations and post-processing analyses used in this work were performed with resources provided by the University of Chicago's Research Computing Center. 

\end{acknowledgments}

\pagebreak

\bibliography{refs}
\bibliographystyle{apalike}

\appendix

\begin{figure*}
    \centering
    \begin{minipage}[t]{0.49\textwidth}
        \centering
        \includegraphics[width=\linewidth]{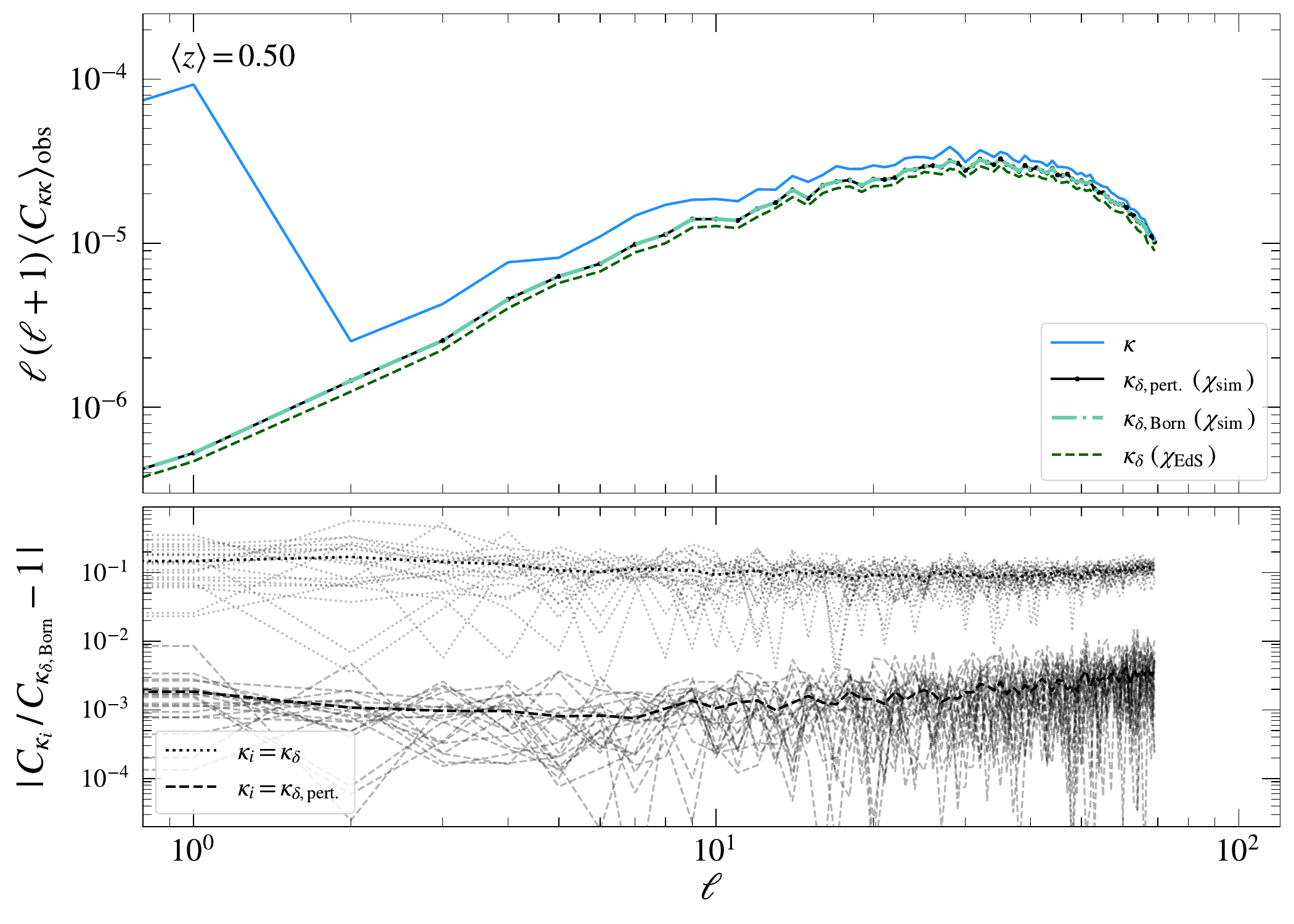}
    \end{minipage}
    \begin{minipage}[t]{0.49\textwidth}
        \centering
        \includegraphics[width=\linewidth]{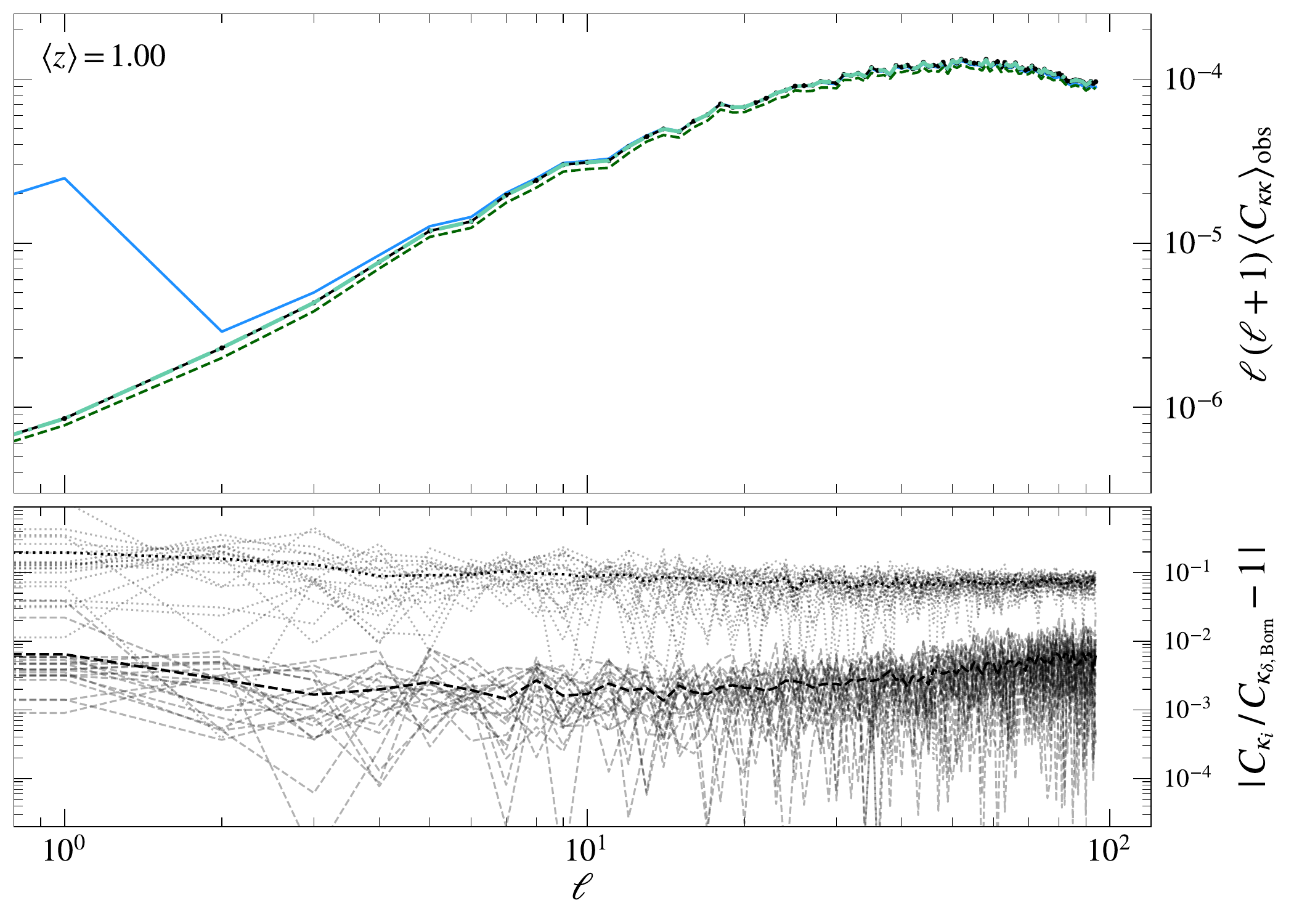}
    \end{minipage}
    \caption{Top: Angular power spectra of nonlinear $\kappa$ (solid curve) and $\kappa_\delta$ calculated along different paths (broken curves). We show the fiducial $\kappa_\delta$ with green dashed curves (using EdS comoving distances), that calculated via simulation comoving distances on perturbed paths ($\kappa_{\delta,{\rm pert.}}$; black curve with points), and with simulation comoving distances on Born paths ($\kappa_{\delta,{\rm Born}}$; light green dot-dashed curve). Bottom: difference in power spectra with respect to the Born approximation for the fiducial $\kappa_\delta$ (dotted) and the perturbed path calculation (dashed) for all 20 observers (faint curves) and the mean (thick curves). Left panels are for $\langle z \rangle = 0.5$ and right for $\langle z \rangle =1.0$.
    Dashed curves in the lower panels indicate the level of validity of the Born approximation in the simulation.
    }
    \label{fig:chi_Born_z0p5_z1p0}
\end{figure*}

\section{Calculating the linearised convergence signals}\label{appx:rel_calcs}

\subsection{Defining the background}

As described in the main text, we must choose a background cosmology for our non-perturbative NR simulations in order to extract the linearised convergence signals. We choose the EdS model used to set the initial data, as detailed in Section~\ref{sec:ics}, i.e. we assume this form of the metric is valid at all times in the simulation. Under this assumption, and alongside our choice of gauge \citet[detailed in][]{Macpherson:2019}, the coordinate time of the simulation hypersurfaces, $t$, is the conformal time, $\eta$. 
In what follows we will give specifics on how we calculate the parameters of the background model from our simulations.

A subtle point is that we cannot simply use $H_0=70$ km/s/Mpc as implied by our choice of $h=0.7$ in the initial data. Due to the finite time resolution of the simulation, we do not land \textit{exactly} on the time slice corresponding to $a=1$ and thus we have $a(z=0) = a_0 \approx 1 + \mathcal{O}(10^{-4})$
which must be accounted for. 
We use the coordinate time of the $z=0$ hypersurfaces (on which the observers live) to calculate the adjusted $H_0 \approx 70 + \mathcal{O}(10^{-2})$ km/s/Mpc which is then used to calculate $\kappa_\delta$ from \eqref{eq:kappaapproxdef}. For both $H_0$ and $H(z)$, we use the EdS relation for the proper-time Hubble parameter $H = 2 /(a\eta)$, where for the latter the time $\eta$ aligns with the constant-time lightcone slices used to calculate the convergence. This calculation of the Hubble parameter differs from that based on observed redshift: $H(z)=H_0 (1+z_s)^{3/2}$ by <~0.5\% for all redshifts. 

To calculate $\bar{\rho}_0$, consequently used to calculate $\delta$, we use a simple Euclidean spatial average over the simulation hypersurface with coordinate time equal to that of the relevant light cone sphere. We set $\Omega_m=1$ and the scale factor is calculated from the \textit{observed} redshift $z_s$ as output from the ray tracer: $a=a_0/(1+z_s)$.

\subsection{Comoving distance and the Born approximation}\label{appx:chi_Born}

In calculating the linearised convergences, an important ingredient is the comoving distance, $\chi$. In the EdS model this is given by
\begin{equation}\label{eq:EdSchi}
    \chi(z_s) = \frac{2}{H_0(1+z_s)} \left(1+z_s - \sqrt{1+z_s}\right),
\end{equation}
however we could also directly calculate the comoving distance in the simulation using the coordinate position of the geodesic at each point. Nominally the $\chi$ that appears in the derivation of the linearised contributions \eqref{eq:kappaapproxdef} and \eqref{eqs:grkappas} is the background comoving distance, so we adopt \eqref{eq:EdSchi} in our main results. Using this relation neglects the deflections of the geodesic along its trajectory; known at the Born approximation. However, the $\delta$ we use in the integral in $\kappa_\delta$ is interpolated to the fully perturbed paths of each geodesic. 
In this appendix, we explore the validity of the Born approximation as well as the validity of \eqref{eq:EdSchi} in relation to the simulation comoving distance.

We calculate the comoving distance in the simulation, $\chi_{\rm sim}$, from the $x^i=(x,y,z)$ position of each point along the geodesic using a Euclidean distance\footnote{The use of a Euclidean distance here is motivated by the fact our chosen background has a flat spatial geometry (including the effect of $\phi$ in these paths makes a negligible impact).} of $\chi_{\rm sim}(t,{\bf n}) = \int_{t_0}^t dt / a(t) = \int_{t_0}^t \sqrt{{\rm d}x^2 + {\rm d}y^2 + {\rm d}z^2}$ where the second equality holds for the path of null rays with ${\rm d}s=0$. Here, ${\rm d}x^i \equiv x^i(t,{\bf n}) - x^i(t_p,{\bf n})$ are the distance traveled by the ray between subsequent time slices ($t$ and $t_p$) in each spatial dimension and each line of sight (${\bf n}$).

We will assess three different methods in calculating $\kappa_\delta$: 1) our fiducial case where we use the EdS $\chi$ from \eqref{eq:EdSchi} and $\delta$ on perturbed paths ($\kappa_\delta$), 2) using both the simulation comoving distance, $\chi_{\rm sim}$, and $\delta$ calculated on fully perturbed paths ($\kappa_{\delta,{\rm pert.}}$), and 3) using both the simulation comoving distance and $\delta$ calculated on Born approximate paths. 

The Born paths in the simulation are calculated using a modified version of the \texttt{mescaline} ray tracer. This works in the same way as the main code, except it uses \textit{only} the initial photon 4--momentum $k^\mu$ for each geodesic to advance the position for all times. Specifically, the ray position is updated according to
\begin{equation}
    \frac{dx_B^i}{dt}(t_{n+1}) = x_B^i(t_{n}) + \Delta t\, \frac{k^i_{\rm ini}}{k^0_{\rm ini}},
\end{equation}
where $k^\mu_{\rm ini}$ is unchanging with time (whereas for the perturbed path $k^\mu$ is updated via the geodesic equation) and set only by the initial direction of the ray (for each \texttt{HEALPix} pixel) and observer 4--velocity \citep[see Section~6.3 of][for more details on the setting of initial data]{Macpherson2023}. The resulting Born positions $x_B^i$ are used to interpolate the density field to obtain $\delta$ along the Born paths as well as to calculate $\chi_{\rm sim}$ as described above. 

The top panel of Figure~\ref{fig:chi_Born_z0p5_z1p0} shows angular power spectra (averaged over 20 observers) for our three methods of calculating $\kappa_\delta$ (as indicated in the legend) as well as the nonlinear convergence, $\kappa$, for reference. Bottom panels show the difference in angular power spectra relative to the Born calculation for fully perturbed paths ($\kappa_{\delta,{\rm pert.}}$; dashed) and for EdS paths ($\kappa_\delta$; dotted). Faint curves here are for individual observers and thicker curves are the average over all 20 observers. The dashed curves show us that Born paths using simulation comoving distances only differ from the fully perturbed paths by 0.1--1\% at all angular scales for both redshift slices. This is mostly consistent with previous theory and Newtonian simulation studies on the validity of the Born approximation \citep{Ferlito:2024,Hilbert:2020,Pratten:2016,Fabbian:2018}.

However, the dotted curves in the lower panels of Figure~\ref{fig:chi_Born_z0p5_z1p0} show us that using the EdS distances makes quite a difference in the resulting $\kappa_\delta$ power spectrum: a consistent $\sim$10\% difference across all angular scales for both redshift slices. We note that this is still slightly smaller than the difference we find between $\kappa_\delta$ and $\kappa$ (dotted curves in the lower panels of Figure~\ref{fig:Cl_comp_all}). However, if we incorporate this throughout all calculations of the linearised convergence signals (also for Doppler, SW, and ISW), we find it improves the match to the nonlinear convergence by up to an order of magnitude (at some angular scales). Regardless of this, we use the EdS distances in the main text since this can be calculated from observables whereas the simulation comoving distance cannot.

\begin{figure*}
    \centering
    \begin{minipage}[t]{0.49\textwidth}
        \centering
        \includegraphics[width=\linewidth]{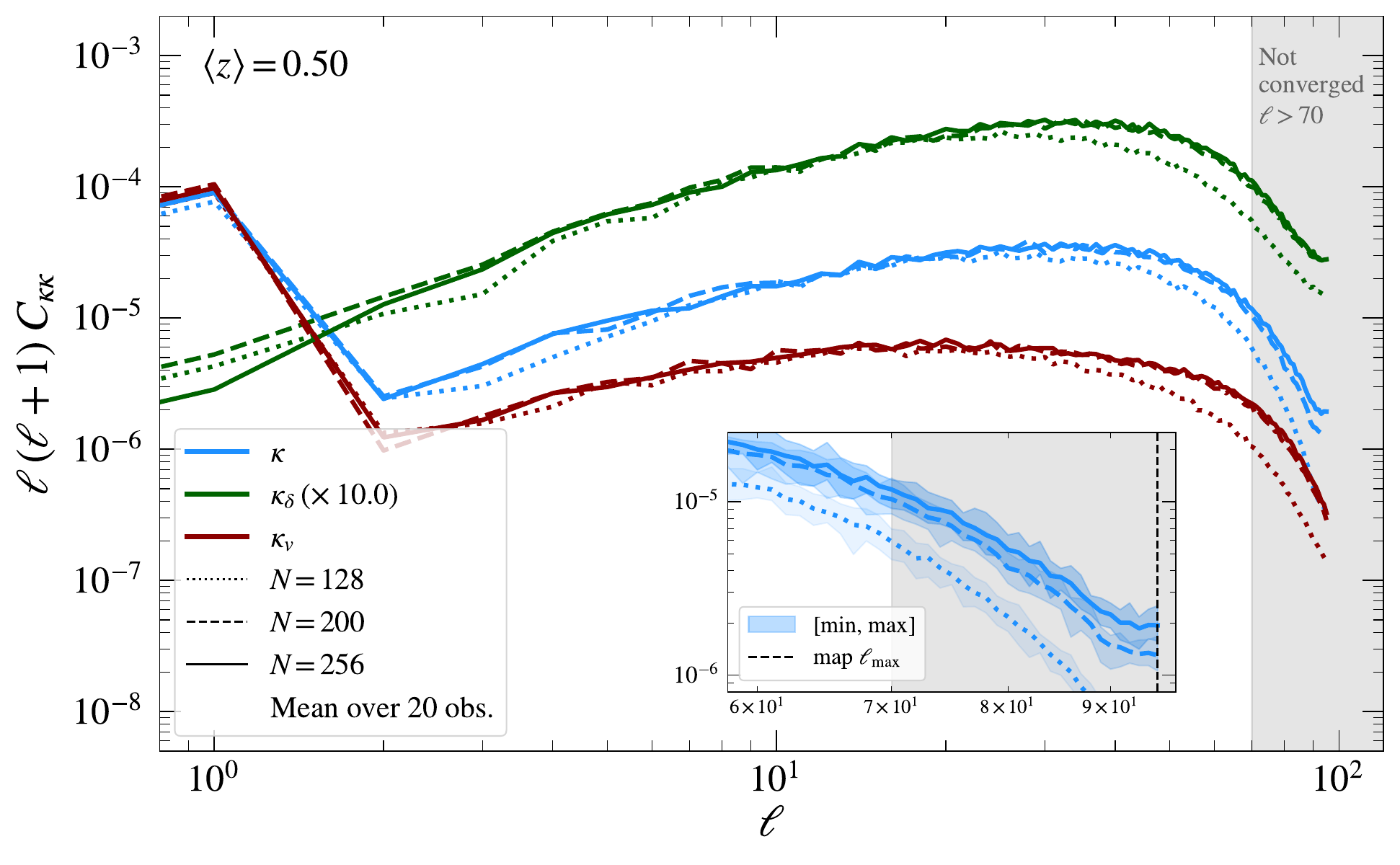}
    \end{minipage}
    \begin{minipage}[t]{0.49\textwidth}
        \centering
        \includegraphics[width=\linewidth]{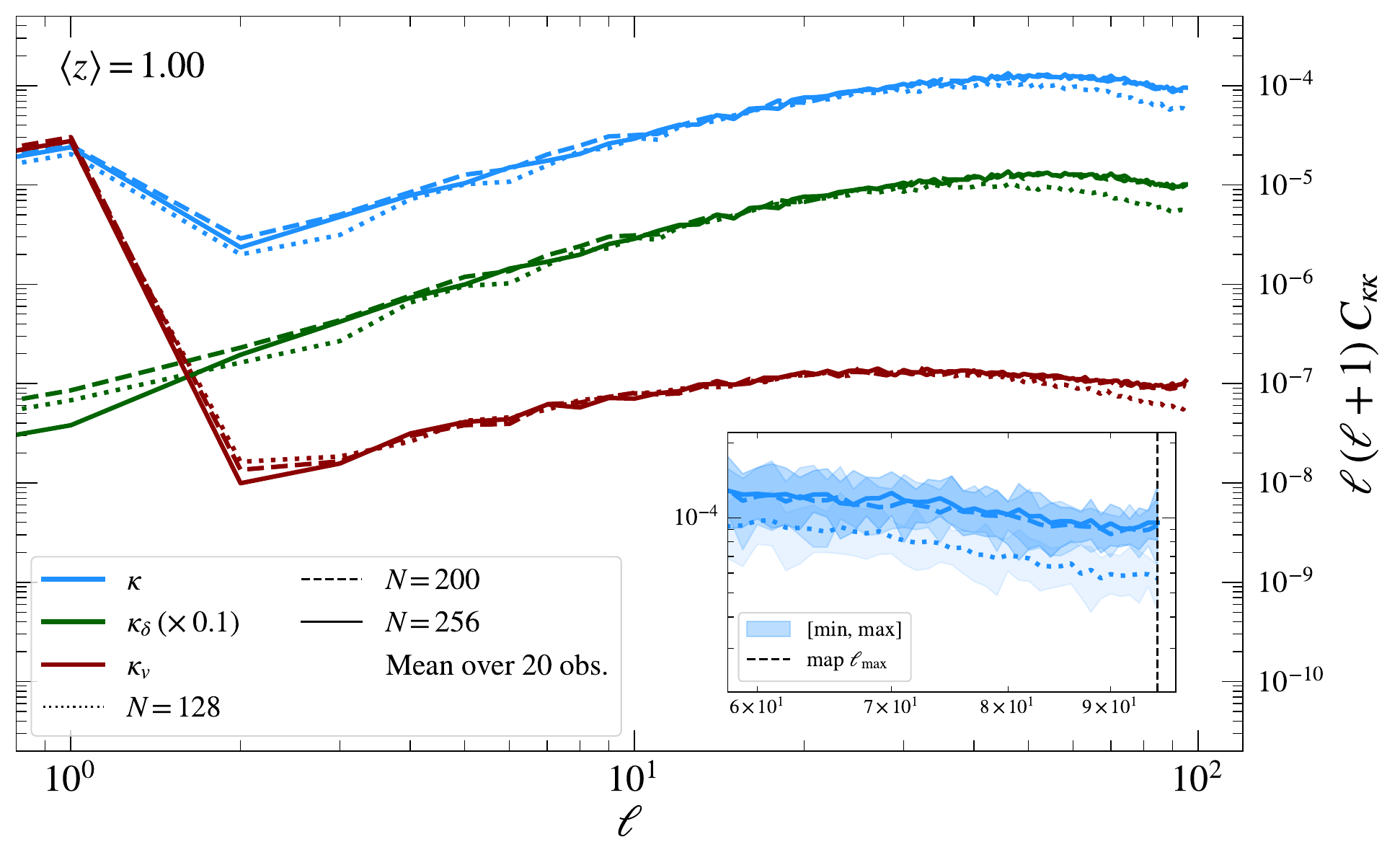} 
    \end{minipage}
    \caption{Numerical resolution study for the convergence angular power spectra averaged over 20 observers for $\langle z \rangle=0.5$ (left) and $\langle z \rangle = 1.0$ (right). Solid curves show $N=256$, dashed curves $N=200$, and dotted curves $N=128$. Blue curves show the nonlinear convergence $\kappa$, green the linearised convergence from the density field $\kappa_\delta$, and red the Doppler lensing $\kappa_v$. Note that $\kappa_\delta$ power spectra is multiplied by a factor (as indicated in each legend) to distinguish it from $\kappa$. The vertical grey shaded region shows the scales which we consider to be not converged. The inset in each panel shows $\kappa$ including the range across all observers shaded for each resolution up to the maximum converged $\ell$ for each redshift.}
    \label{fig:Cl_restest_z0p5_z1}
\end{figure*}

\subsection{Relativistic contributions}

To calculate the relativistic contributions to the linear convergence in \eqref{eqs:grkappas} we must assume the metric in our NR simulations remains of the form \eqref{eq:pertmetric} beyond the initial slice. This means we are neglecting vector and tensor contributions, which is common since these are expected to be subdominant with respect to the scalar perturbation in a spatial sense \citep{Adamek:2016a,Lu:2009} as well as in their effect on the convergence itself \citep{Andrianomena:2014}.  
Under these assumptions, the determinant of the spatial metric is
\begin{equation}\label{eq:phi}
    \gamma = a(t)^6 (1-2\phi)^3,
\end{equation}
which we use to extract $\phi$ from the $\gamma_{ij}$ data that is output from the ET. 

The peculiar velocity ${\bf v}$ in the Doppler term \eqref{eq:kappadop} is different to the velocity output from the ET, with the former defined as $v^i=dx^i/dt$ and the latter $v_{\rm ET}^i=u^i/u^0$. Thus, we have $v^i=\alpha \, v_{\rm ET}^i$ where $\alpha$ is the lapse function.

We also need $\phi '$ for the ISW contribution, which we extract from the trace of the extrinsic curvature, $K$. For the linearly-perturbed metric \eqref{eq:pertmetric}, we have
\begin{equation}\label{eq:phidash}
    K = - \frac{3 \mathcal{H}(t)}{\alp} + \frac{3 \phi '}{\alp\, (1-2\phi)},
\end{equation}
where we use $K$ and $\alp$ output from the ET and subsequently interpolated to the position of the geodesic. %
We find $\phi '< 10^{-8}$ and $\phi \sim 10^{-5}$--$10^{-4}$ for all redshifts (see Figure~\ref{fig:phi_derivs_v_z} as described in Appendix~\ref{appx:kappa_terms}).

All simulation outputs (i.e., $v^i_{\rm ET}$, $\rho_0$ for $\delta$, $\gamma$, and $K$) used to calculate the perturbations are interpolated to the position of the geodesic from the 3--dimensional simulation output using a cubic spline interpolator via a modified version of the \texttt{mescaline} ray tracer.

Importantly, the $a(t)$ and $\mathcal{H}(t)$ used in \eqref{eq:phi} and \eqref{eq:phidash} must be calculated according to qualities of the spatial hypersurfaces themselves, rather than from the observed redshifts as in the previous section. %
We use the conformal-time relations $\mathcal{H}(t)=2/t$ and $a(t)=a_{\rm ini}(t/t_{\rm ini})^2$, where $a_{\rm ini}$ and $t_{\rm ini}$ are the initial values at the beginning of the simulation. We remind the reader that in our chosen gauge \citep[see][for details]{Macpherson:2019} the simulation coordinate time, $t$, is coincident with the conformal time so long as we are close to the initial metric \eqref{eq:pertmetric}.

\begin{figure}
    \centering
    \begin{minipage}[t]{0.49\textwidth}
        \centering
        \includegraphics[width=\linewidth]{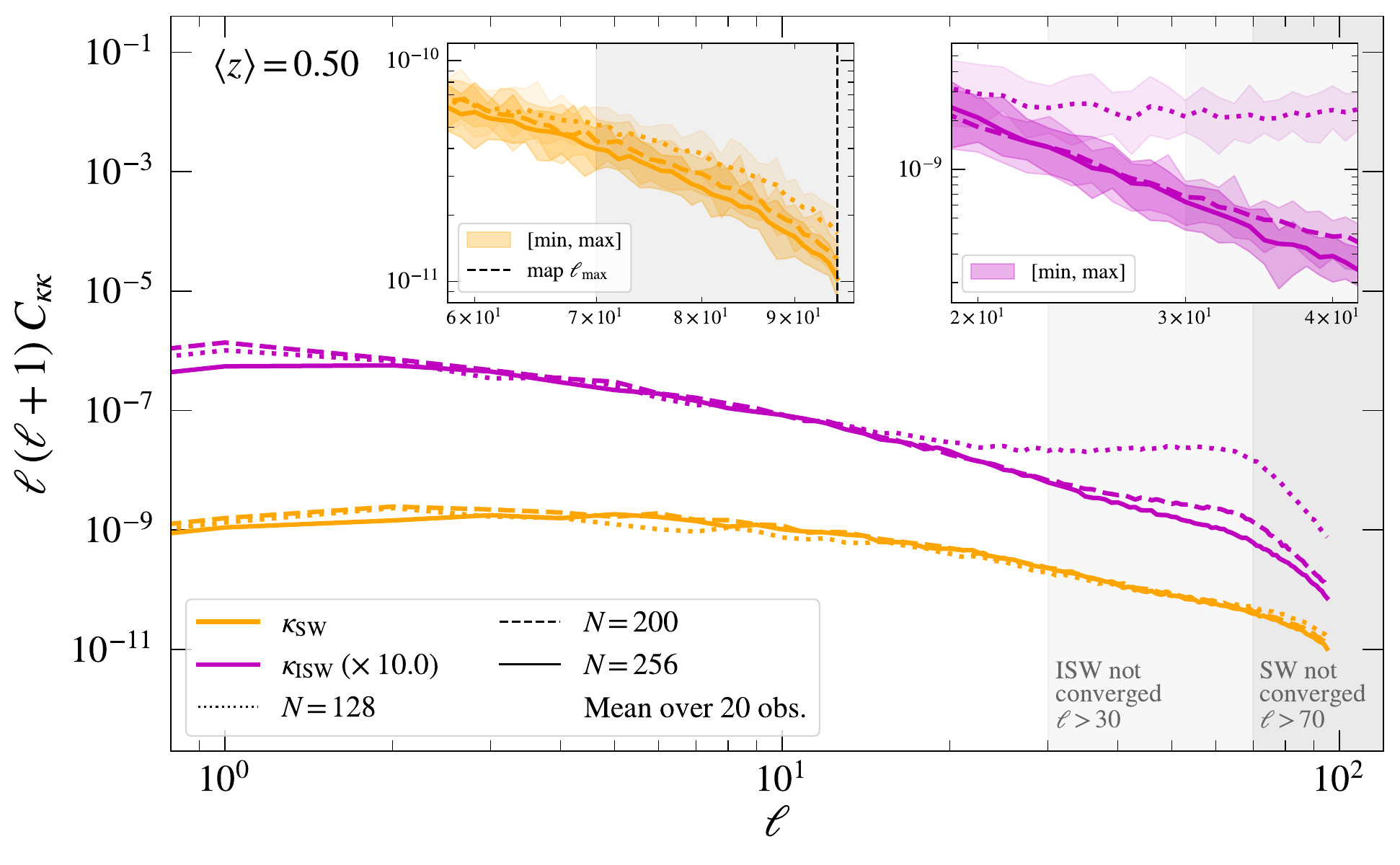}
    \end{minipage}
    \begin{minipage}[t]{0.49\textwidth}
        \centering
        \includegraphics[width=\linewidth]{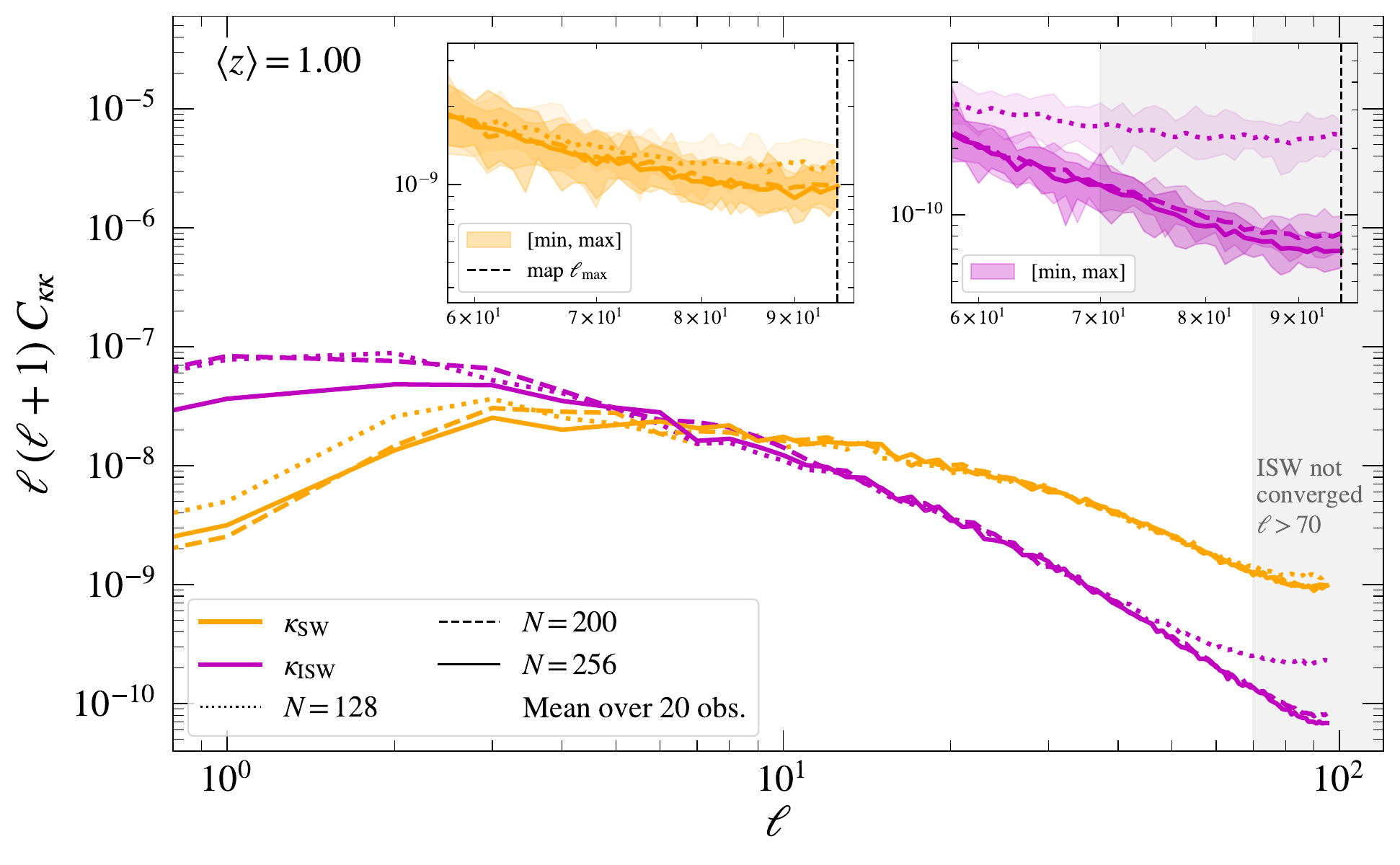} 
    \end{minipage}
    \caption{Numerical resolution study for convergence of the SW and ISW power spectra averaged over 20 observers for $\langle z \rangle = 0.5$ (top panel) and $\langle z \rangle = 1.0$ (bottom panel). Magenta curves show the ISW signal and the gold curves show the SW signal, with dotted, dashed, and solid curves indicating resolutions $N=128,200$ and 256, respectively. The insets show zoom-ins on each of the SW (left inset) and ISW (right inset) with shaded regions around each curve showing the spread from minimum to maximum across all observers. Grey vertical shaded regions show the zones we consider not converged for each redshift and signal, as labeled in each case.}
    \label{fig:Cl_restest_sws_z0p5_z1}
\end{figure}

\section{Robustness of results}

In this Appendix, we perform robustness tests to ensure our results remain consistent under changes to the numerical resolution in Appendix~\ref{appx:convergence}, and we study the level of constraint violation in the simulations in Appendix~\ref{appx:constraints}.

\subsection{Numerical convergence}\label{appx:convergence}

First we repeat our main results in a set of three simulations with different numerical resolution of: $N=128, 200$, and 256.
The initial data for these simulations was drawn from the same power spectrum (with different random seeds) sampling the same physical scales, and all have the same box size of $L=3072\,h^{-1}$ Mpc. 
As mentioned in the main text, we require a minimum scale cut above the simulation grid resolution in order to minimise the level of numerical noise associated with under-sampling the small-scale modes. For our fiducial simulation with $N=256$, the minimum scale cut is $10\Delta x$. To keep the minimum \textit{physical} scale (of $\sim 120\,h^{-1}$ Mpc) the same while changing the numerical resolution, this requires a minimum cut for $N=128$ of $5\Delta x$ and for $N=200$ of $7.8125\Delta x$. During the simulation, structures below this scale do form and this choice only restricts the initial data; which in general means that the higher-resolution simulations will have a higher degree of nonlinearity simply because there is resolution at smaller scales. 
This means we cannot perform a Richardson extrapolation to determine the error bars on our calculations, since this requires the physical scenario to remain identical to isolate the change in resolution. 
In our case, this would require the simulations to be linear in nature \citep[see e.g. Appendix~C1 of][]{Macpherson:2021} since nonlinearity will inherently change the structures when smaller modes are resolvable. 
We thus assess numerical convergence in a \textit{statistical} sense. Specifically, for each resolution simulation we calculate the angular power spectrum of the lensing convergence for a set of 20 observers. Due to this limited number of samples, we do not expect them to match exactly. We will consider our results numerically converged if the mean across all observers, as well as the spread of the distribution, is roughly consistent across resolutions.

Figure~\ref{fig:Cl_restest_z0p5_z1} shows the convergence angular power spectra for $\kappa$ (blue), $\kappa_\delta$ (green) and $\kappa_v$ (red) at three resolutions $N=256$ (solid), $N=200$ (dashed), and $N=128$ (dotted) for $\langle z \rangle=0.5$ (left panel) and $\langle z \rangle = 1.0$ (right panel). Each power spectra is averaged over 20 observers in each simulation. Note that we have multiplied the $\kappa_\delta$ power spectra by a factor of 10 (0.1) to be able to clearly distinguish it from $\kappa$ in the left (right) panel. 
The vertical shaded region in the left panel shows the angular scales for which we consider our results \textit{not} numerically converged. The inset shows a zoom-in of the transition from converged to not-converged for the $\kappa$ power spectrum. 
Here we see that $\ell\approx 70$ represents roughly the point above which the distributions over observers (shaded regions) begin to diverge from one another for $N=200$ (dashed) and $N=256$ (solid). In the right panel, also shown in the inset, all angular scales up to the resolution of the map ($\ell=95$) are unchanged for $N=200$ to $N=256$ and thus are numerically converged. The story is similar for $\kappa_\delta$ and $\kappa_v$ for both redshift cases. 

Figure~\ref{fig:Cl_restest_sws_z0p5_z1} shows the numerical convergence of the SW (gold) and ISW (magenta) angular power spectra for the same three resolutions. Top panel shows $\langle z \rangle = 0.5$ and the bottom panel shows $\langle z \rangle = 1.0$. For each plot, we show two insets of the SW and ISW including the spread from minimum to maximum across observers shown as a shaded region for each resolution. The vertical grey bands show the non-convergence regions we determine (as labeled): for the top panel the SW signal is converged up to $\ell\approx 70$ (as we found for the other contributions in Figure~\ref{fig:Cl_restest_z0p5_z1}) and we find the ISW signal is only converged up to $\ell\approx 30$. For the bottom panel, the SW signal is converged to $\ell_{\rm max}$ and ISW up to $\ell\approx 70$. 

\begin{figure*}
    \centering
    \begin{minipage}[t]{0.49\textwidth}
        \centering
        \includegraphics[width=\linewidth]{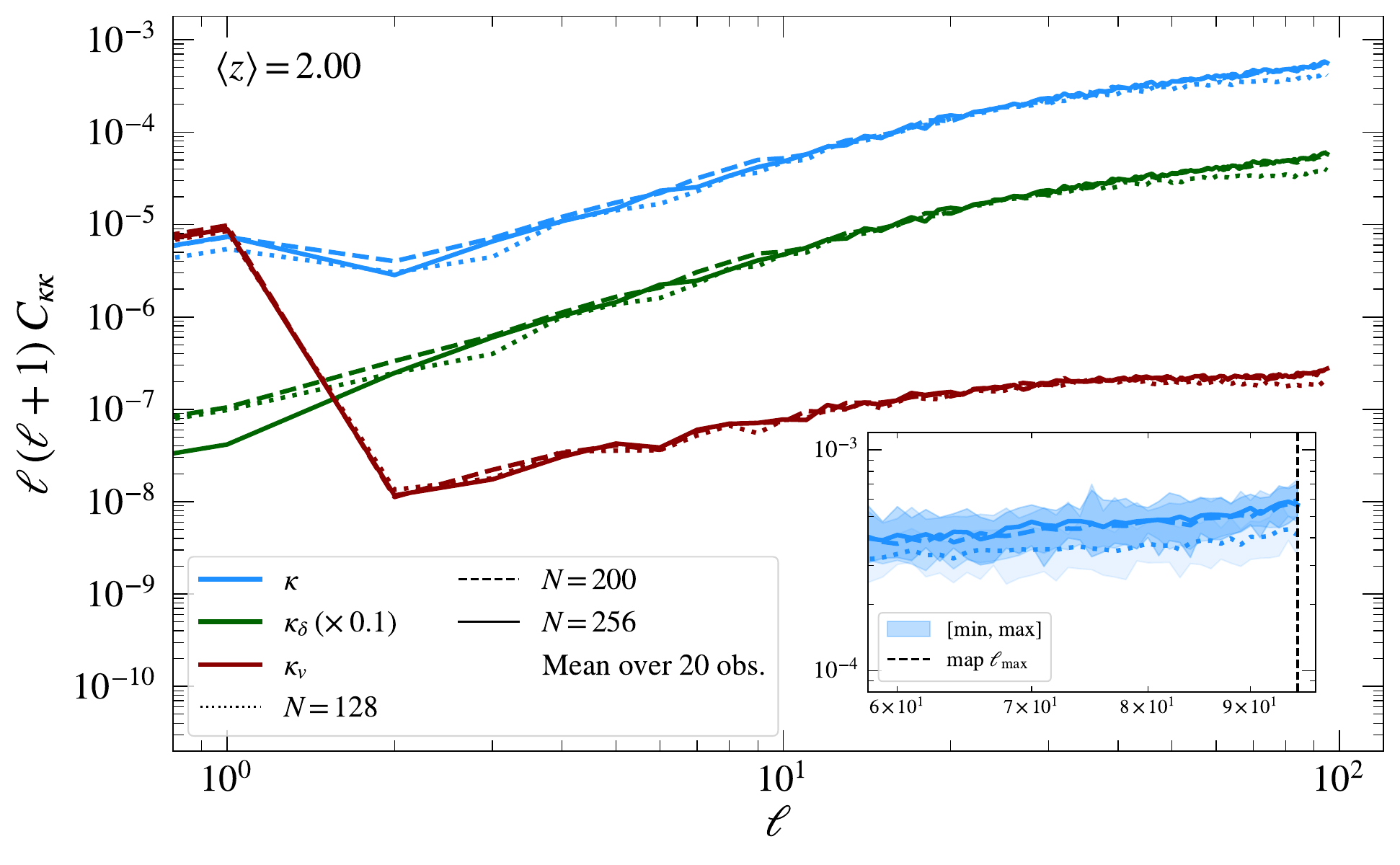}
    \end{minipage}
    \begin{minipage}[t]{0.49\textwidth}
        \centering
        \includegraphics[width=\linewidth]{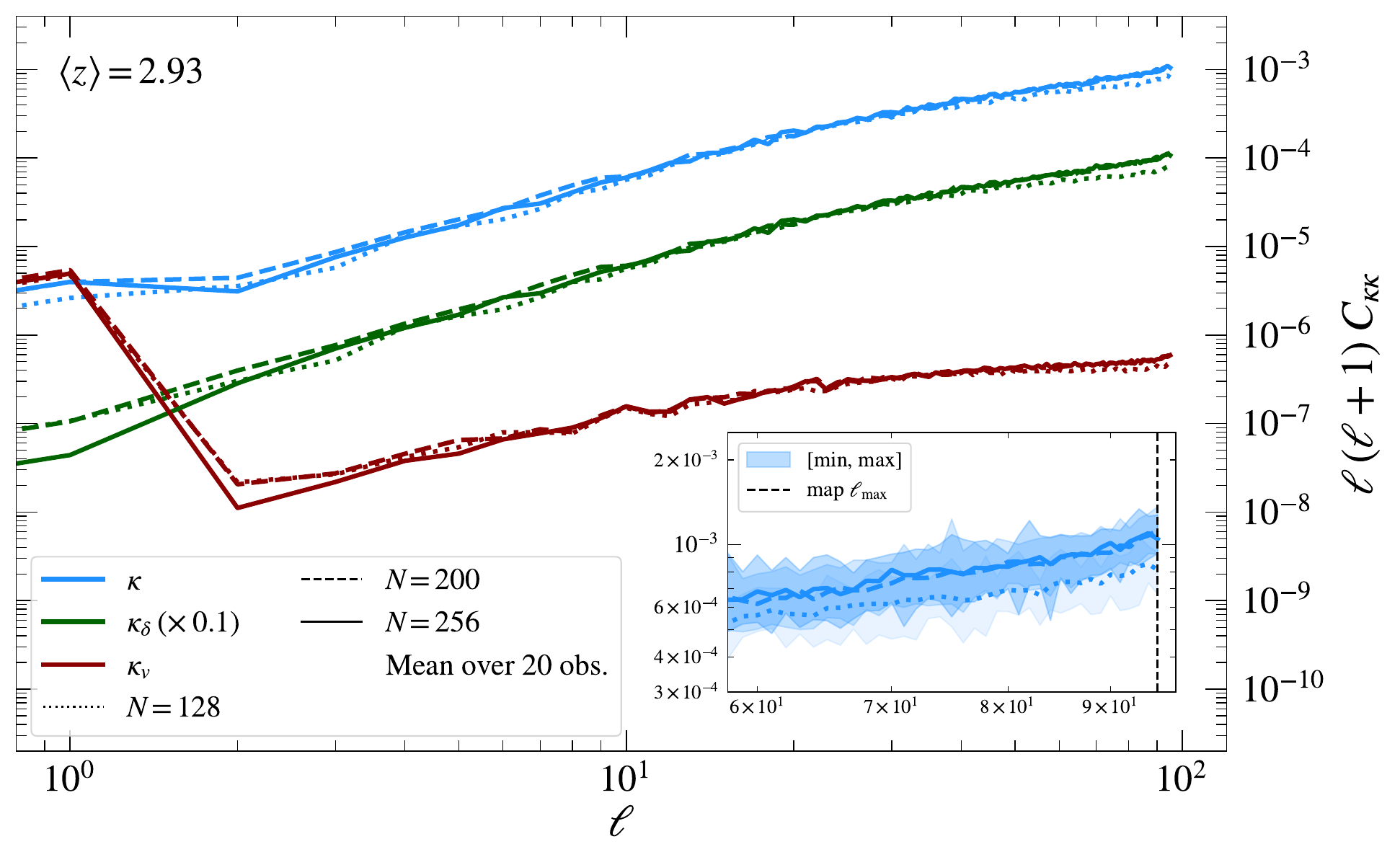} 
    \end{minipage}
    \caption{Numerical resolution study for the angular power spectra for $\langle z \rangle = 2.0$ (left) and $\langle z \rangle = 2.93$ (right). Colours, line styles, and shading are the same as in Figure~\ref{fig:Cl_restest_z0p5_z1}.}
    \label{fig:Cl_restest_z2_z3}
\end{figure*}
Figure~\ref{fig:Cl_restest_z2_z3} shows the numerical convergence, as in Figure~\ref{fig:Cl_restest_z0p5_z1}, for $\langle z \rangle = 2.0$ (left panel) and $\langle z \rangle = 2.93$ (right panel). We find numerical convergence for the angular power spectra up to the maximum resolution of the map. We show these higher redshift slices to show that our results remain converged for the entire range shown in Figure~\ref{fig:Cldiff_vs_z} (the difference in the power spectrum vs redshift).
We find similar convergence patterns up to $\ell\sim \ell_{\rm max}$ for the SW and ISW power spectra for these higher redshifts.

\subsubsection{Summary of numerical convergence study}
We find that all angular power spectra except ISW are not converged for $\ell>70$ for $z\lesssim$~0.4--0.6. For even lower redshifts of $z=$~0.1--0.3 (which we do not show a plot for) we find our results are not converged for $\ell\gtrsim$~50--60. For all $z>0.6$ we find convergence to $\ell_{\rm max}=95$. 
For the ISW contribution, the signal is so small in magnitude for $z\lesssim 0.1$ that we do not find convergence past the first few multipoles. For $z=$~0.1--0.3, the signal converges until $\ell\sim 10$. Once we reach $z=$~0.5--0.8, the ISW signal converges up until $\ell=30$, and for $z=1$ we find convergence until $\ell=70$. Beyond $z=2$, ISW also converges up to $\ell_{\rm max}$. Importantly, all plots in the main text \textit{only} show results for which we find numerical convergence.

\subsection{Constraint violation}\label{appx:constraints}
\begin{figure*}
    \centering
    \includegraphics[width=\linewidth]{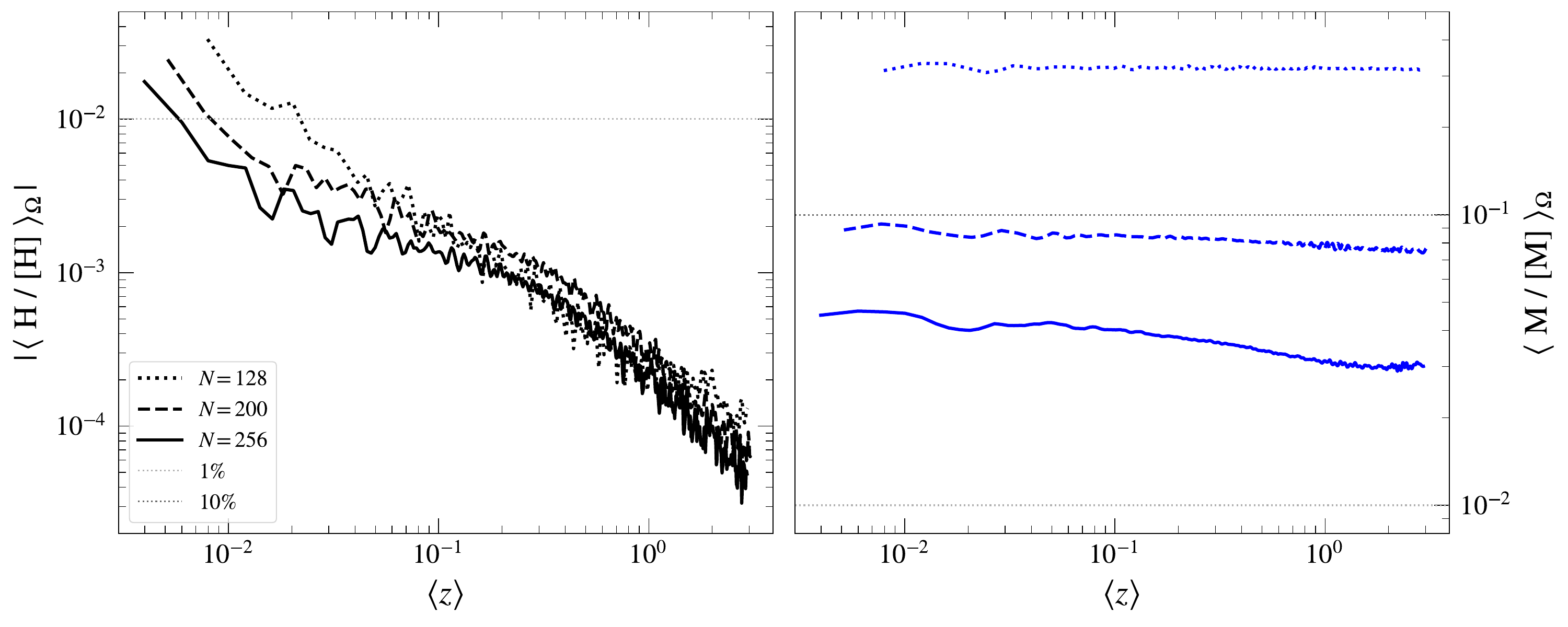}
    \caption{Mean constraint violation on the sky for three resolutions $N=128$ (dotted), $N=200$ (dashed), and $N=256$ (solid) for both the Hamiltonian (left panel) and momentum (right panel) constraints. Curves show the mean across 20 observers in each simulation as a function of mean redshift of the slice. }
    \label{fig:HM_avgs}
\end{figure*}

The Hamiltonian and momentum constraints \eqref{eqs:constraints} are identically zero for an exact solution of Einstein's equations, however, will be slightly non zero due to numerical error. This error arises from an initial violation in the constraints when generating the initial data (i.e. if one uses approximations) as well as the accumulation of error due to the finite accuracy of the numerical evolution scheme. The former, as described in Section~\ref{sec:ics}, is determined by the fourth-order accuracy of the \texttt{mescaline} finite-difference derivatives used to calculate, e.g., the Ricci tensor. The accumulation of error in the constraint violation is determined by the numerical accuracy of the ET simulations themselves, which is also fourth order. However, since this accumulates with each timestep we expect this to be the dominant source of violation of the constraints. 

In any case, the violation we expect is always proportional to the grid spacing to the power of four, i.e. $(\Delta x)^4$. Thus, the violation should decrease as we increase the numerical resolution if it is indeed dominated by numerical error. 
In this Appendix, we assess the level of violation in the constraints for the same simulations we studied in Appendix~\ref{appx:convergence} with three resolutions $N=128, 200$, and 256. \texttt{mescaline} calculates the normalised constraint violation $H/[H]$ and $M/[M]$ where $M\equiv \sqrt{M_i M^i}$ ($H$ and $M_i$ are defined in \eqref{eqs:constraints}) and the normalisations are determined by a sum of squares of the contributing terms, namely
\begin{subequations}
    \begin{align}
        [H] &\equiv \sqrt{ \mathcal{R}^2 + (K_{ij}K^{ij})^2 + (K^2)^2 + (16\pi\rho)^2 }, \\
        [M] &\equiv \sqrt{ D_j K^j_{\phantom{j}i} D_l K^{li} + D_i(K) D^i(K) + (8\pi)^2 S_i S^i}.
    \end{align}
\end{subequations}
These normalised violations give us a dimensionless measure of the level of constraint violation in the simulation \citep[see also, e.g.][]{Giblin:2017b,Macpherson:2019}.
We use the \texttt{mescaline} ray tracer to output the value of $H/[H]$ and $M/[M]$ at every position along each geodesic for all 20 observers in all three simulations. 

Figure~\ref{fig:HM_avgs} shows the mean across the sky of the Hamiltonian (left panel; absolute value) and momentum (right panel) violations, also averaged across all 20 observers. Solid curves show $N=256$, dashed for $N=200$, and dotted for $N=128$. Faint dotted horizontal lines show 1\% and 10\% violations, for reference. The magnitude of the mean violation is $\lesssim$~1--4\% for both constraints for our highest resolution (fiducial results) and, importantly, the violation reduces as the resolution is increased.

\begin{figure*}
    \centering
    \includegraphics[width=\linewidth]{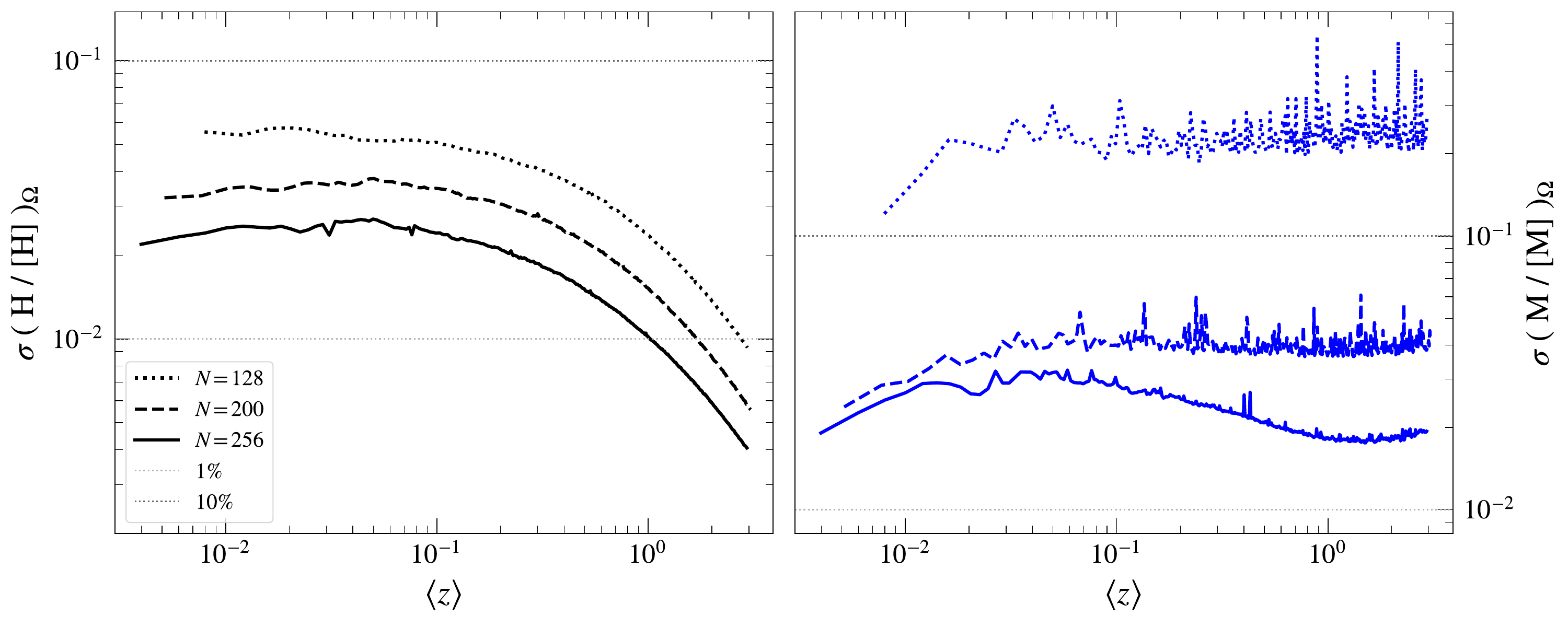}
    \caption{Standard deviation of the constraint violation across the sky for the three resolutions as shown in Figure~\ref{fig:HM_avgs}. Here each curve is the mean across 20 observers and shows the typical fluctuations in constraint violation on the sky as a function of mean redshift.}
    \label{fig:HM_stds}
\end{figure*}
Figure~\ref{fig:HM_stds} shows the standard deviation across the sky for the Hamiltonian (left panel) and momentum (right panel) violation, again averaged over all 20 observers. As expected, this is higher in magnitude than the mean values in Figure~\ref{fig:HM_avgs}, since we expect the violation to be higher in regions with larger density contrasts (since under-sampling modes will contribute to this violation). Still, the violation remains $\sim 2$\% for our highest resolution, and reduces as we increase resolution. 

Importantly, as we showed in Appendix~\ref{appx:convergence}, our main results do not change significantly for $N=200\rightarrow 256$ for most redshifts and angular scales. If these results were dominated by constraint violation we would expect them to continue changing as we increased resolution. In conclusion, by combining the convergence studies in Appendix~\ref{appx:convergence} with those presented here we can conclude that our main results are not dominated by finite-difference error and/or constraint violation.

\section{Approximations in deriving $\kappa_\delta$}\label{appx:kappa_terms}

As mentioned in Section~\ref{sec:kappa_approx}, in deriving the linearised convergence from the density field, $\kappa_\delta$ in \eqref{eq:kappaapproxdef}, several terms at first and second order have been neglected. In this Appendix, we will explore the contribution from these various terms and whether they might be a source for the difference we found between the nonlinear and linearised convergence.

The convergence from gravitational lensing is
\begin{equation}\label{eq:kg}
    \kappa_g = \int^{\chi_s}_0 \frac{\chi(\chi_s - \chi)}{\chi_s} \nabla_\perp^2 \phi,
\end{equation}
where $\nabla_\perp$ is the transverse Laplacian in the 2--dimensional screen-space plane (which is perpendicular to the geodesic at all times), which may be written as 
\begin{equation}\label{eq:nablaperp}
    \nabla_\perp^2 = \nabla^2 - ({\bf n} \cdot \nabla)^2 + \frac{2}{\chi} \,{\bf n} \cdot \nabla.
\end{equation}
Here, $\nabla^2$ is the regular spatial Laplacian and $\nabla$ is the vector gradient. 

It is useful to introduce the total derivative along the geodesic (which for our perturbative description can be expressed in terms of the background co-moving distance, $\chi$), as
\begin{align}
    \frac{\rm d}{{\rm d}\chi} &= - \frac{\partial}{\partial \eta} - {\bf n}\cdot \nabla,\label{eq:ddchi} %
\end{align}
which accounts for both temporal ($\eta$; with the negative sign since we are moving backwards in time away from the observer) and spatial ($\nabla$; with the $-{\bf n}$ since we are moving in the direction of observation) changes along the geodesic.
We can use \eqref{eq:ddchi} in combination with the transverse Laplacian \eqref{eq:nablaperp} to write the term in the integrand of \eqref{eq:kg} as 
\begin{equation}\label{eq:nabla_perp_phi}
    \nabla_\perp^2 \phi = \nabla^2 \phi - \frac{{\rm d}^2\phi}{{\rm d}\chi^2} - \frac{\partial}{\partial\eta} \frac{{\rm d}\phi }{{\rm d}\chi} - \frac{{\rm d}\phi' }{{\rm d}\chi} - \phi '' - \frac{2}{\chi} \left( \frac{{\rm d}\phi}{{\rm d}\chi} + \phi ' \right).
\end{equation}
See also Appendix~A of \citet{Bacon:2014}\footnote{Note that equation (A2) in \citet{Bacon:2014} is incorrect and we have corrected this in \eqref{eq:nabla_perp_phi} here.} and Section~5.2.2 of \citet{Fleury:2015a}. 

Usually, all terms after the first one are neglected. Then the Poisson equation can be substituted to relate $\nabla^2\phi$ to the density fluctuation $\delta$ and finally arrive at $\kappa_\delta$ in \eqref{eq:kappaapproxdef}. The Poisson equation is widely adopted as an approximation to relativistic dynamics in the regime of small velocities and gravitational potentials; which are both satisfied in the simulations we use here. Naively, we would expect this relation to hold with very good accuracy. However, we will still calculate the precision of this relation in our simulations in Section~\ref{appx:del2phi} below. 

We would like to calculate \textit{all terms} on the right hand side of \eqref{eq:nabla_perp_phi} and determine their exact size. However, since time derivatives of $\phi$ are expected to be very small in our simulations, we will neglect $\phi ''$ as well as $\frac{\partial}{\partial\eta} \frac{{\rm d}\phi }{{\rm d}\chi}$. We will thus assess the size of the derivatives of $\phi$ along the geodesic with respect to $\chi$---which has not been done before in a fully nonlinear simulation.

\subsection{Size of neglected terms}\label{appx:kgterms}

\subsubsection{$\phi$ and its derivatives}\label{sappx:phiderivs}

We calculate the total derivatives with respect to $\chi$ using the value of $\phi$ along each geodesic output from the ray tracer. We use $\phi$ and $\chi$ calculated along each geodesic, as described in Section~\ref{appx:rel_calcs}, in the \texttt{numpy.gradient} function to calculate ${\rm d}\phi/{\rm d}\chi$ (and a second gradient for ${\rm d}^2\phi/{\rm d}\chi^2$). 

\begin{figure}
    \centering
    \begin{minipage}[t]{\linewidth}
        \centering
        \includegraphics[width=\linewidth]{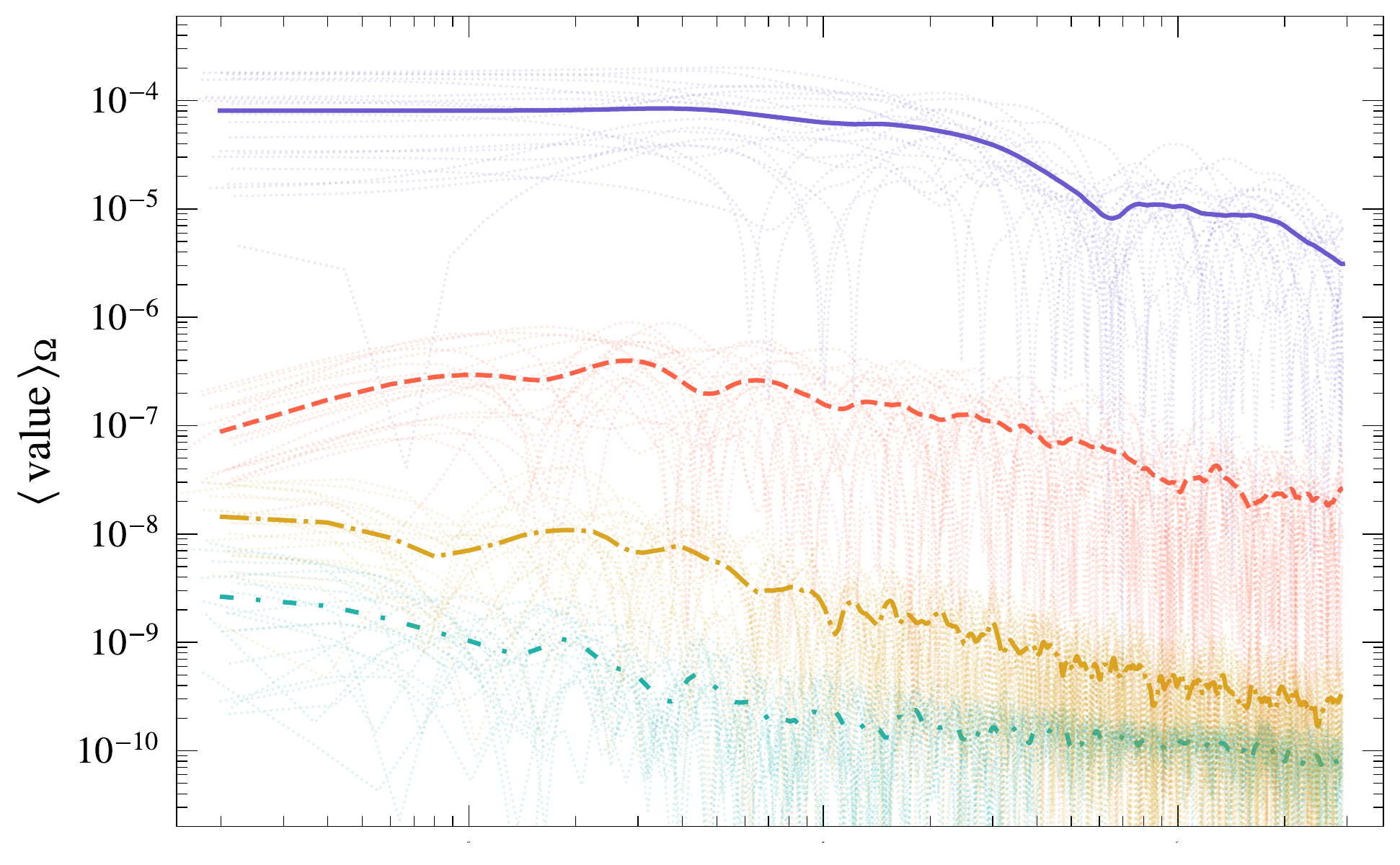}
    \end{minipage}
    \begin{minipage}[t]{\linewidth}
        \centering
        \includegraphics[width=\linewidth]{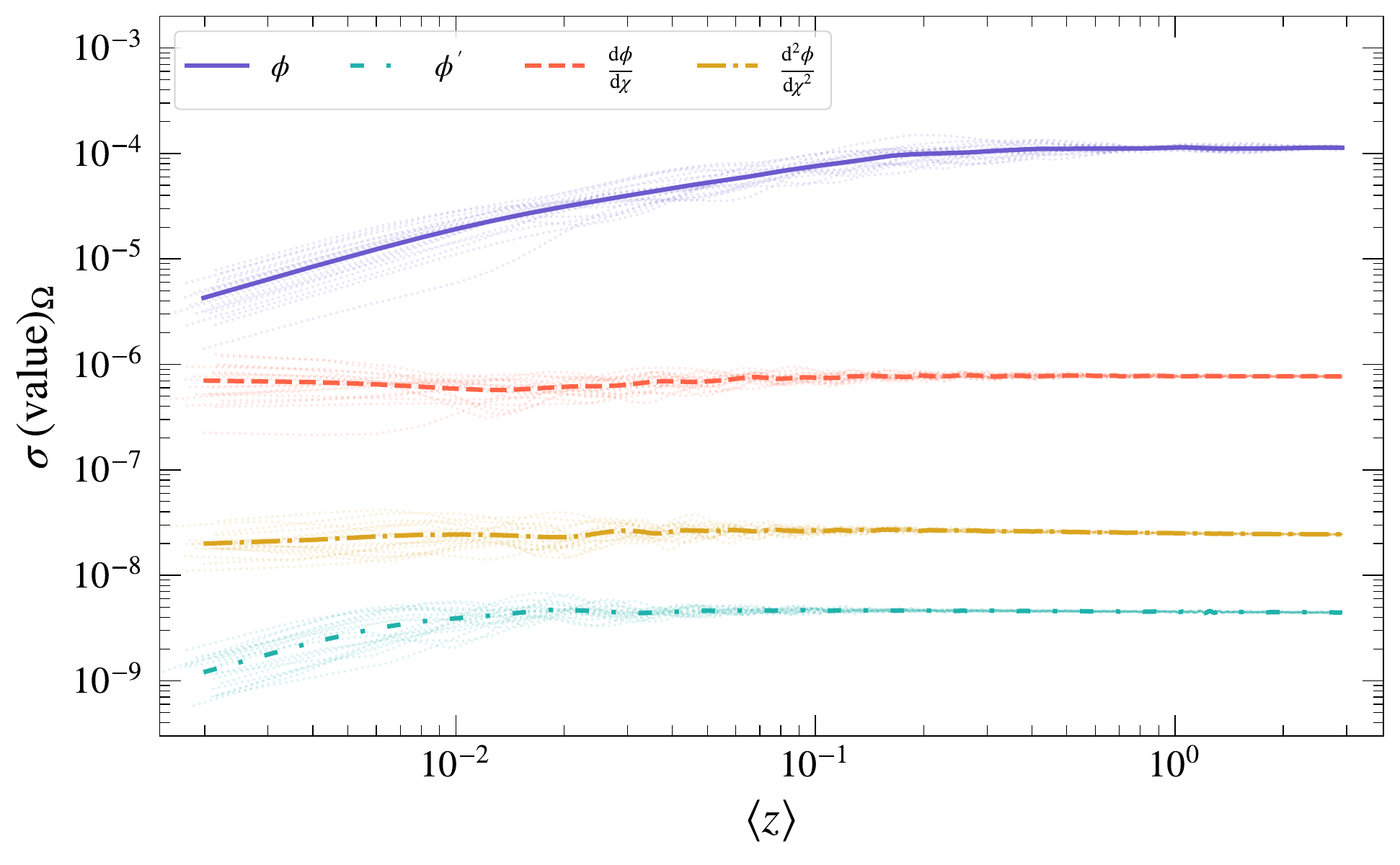}
    \end{minipage}
    \caption{Size of $\phi$ and its derivatives as a function of mean redshift. Top panel shows the mean value of each quantity (as indicated in the legend) on the sky with thick curves indicating the mean over 20 observers and the thin curves for individual observers. Bottom panel shows the standard deviation $\sigma$ of fluctuations across the sky for the same quantities.}
    \label{fig:phi_derivs_v_z} 
\end{figure}
Figure~\ref{fig:phi_derivs_v_z} shows the redshift evolution of $\phi$ (solid blue curve), $\phi '$ (sparse dot-dashed cyan curve), ${\rm d}\phi/{\rm d}\chi$ (dashed red curve), and ${\rm d}^2\phi/{\rm d}\chi^2$ (dot dashed gold curve). Top panel shows the mean value on the sky and bottom panel shows the standard deviation of fluctuations on the sky for that redshift. Faint dotted curves are the values for individual observers and thick curves are the mean over the 20 observers. 
We find that ${\rm d}\phi '/{\rm d}\chi$ is consistent with numerical error (i.e., it reduces proportional to an increase in resolution), so we exclude it from these plots and the rest of this Appendix. All other quantities are robust to changes in resolution (by a similar standard as is discussed in Appendix~\ref{appx:convergence}). 

Figure~\ref{fig:phi_derivs_v_z} gives an interesting quantification of the size of these derivatives as a function of redshift. We emphasise to the reader that these quantities are statistics across the observer's \textit{lightcone slices} and not spatial slices of the simulation. So the evolution may not align exactly with what we expect from, e.g., perturbation theory as calculated on spatial sections at a given time. In general, we find that all derivatives of $\phi$ are subdominant to $\phi$ itself for all redshifts, especially the mean values across the sky which are always at least 2--3 orders of magnitude lower. 

At low redshift, we might be surprised to see that the standard deviation of fluctuations in $\phi$ on the sky reduces by an order of magnitude (solid curve in the bottom panel). The distribution of $\phi$ fluctuations on the sky is not well described by a Gaussian for $z\lesssim 0.25$, so the standard deviation is not necessarily a good measure of the typical fluctuations for these lower redshifts.

\subsubsection{Neglected convergence signals}

Next, we will calculate the convergence signals associated with these terms by writing:
\begin{align}\label{eq:kappas_neglected}
    \kappa_g = \kappa_\delta + \kappa_{{\rm d}^2\phi} + \kappa_{{\rm d}\phi}, %
\end{align}
where the individual contributions are defined as
\begin{subequations}\label{eqs:kextras}
    \begin{align}
        \kappa_{{\rm d}^2\phi} &\equiv - \int^{\chi_s}_0 f(\chi)\; \frac{{\rm d}^2\phi}{{\rm d}\chi^2} \;{\rm d}\chi , \label{eq:kd2phi}\\
        \kappa_{{\rm d}\phi} &\equiv - \int^{\chi_s}_0 f(\chi)\; \frac{2}{\chi} \,\left( \frac{{\rm d}\phi}{{\rm d}\chi} + \phi ' \right) \;{\rm d}\chi, \label{eq:kdphi}
    \end{align}
\end{subequations}    
where $f(\chi)\equiv \chi(\chi_s - \chi)/\chi_s$ and $\kappa_\delta$ is as defined in \eqref{eq:kappaapproxdef}. We have excluded the convergence signal from the term proportional to ${\rm d}\phi'/{\rm d}\chi$ since, as mentioned above, we find it consistent with numerical error. %
We calculate the terms in the integral for each of these as described in Section~\ref{sappx:phiderivs} above and sum along each line of sight. 

\begin{figure*}
    \centering
    \begin{minipage}[t]{0.49\textwidth}
        \centering
        \includegraphics[width=\linewidth]        {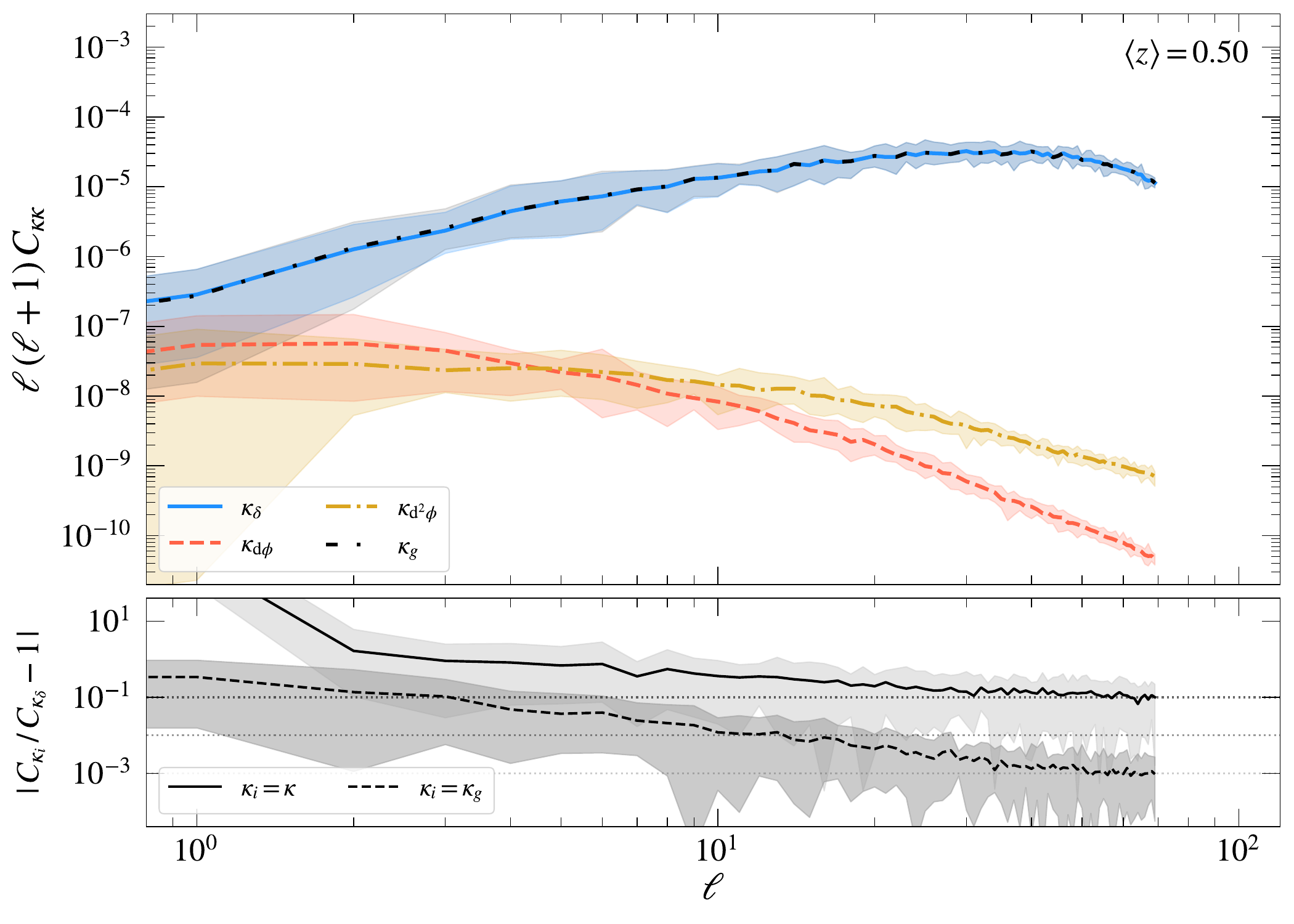}
    \end{minipage}
    \begin{minipage}[t]{0.49\textwidth}
        \centering
        \includegraphics[width=\linewidth]{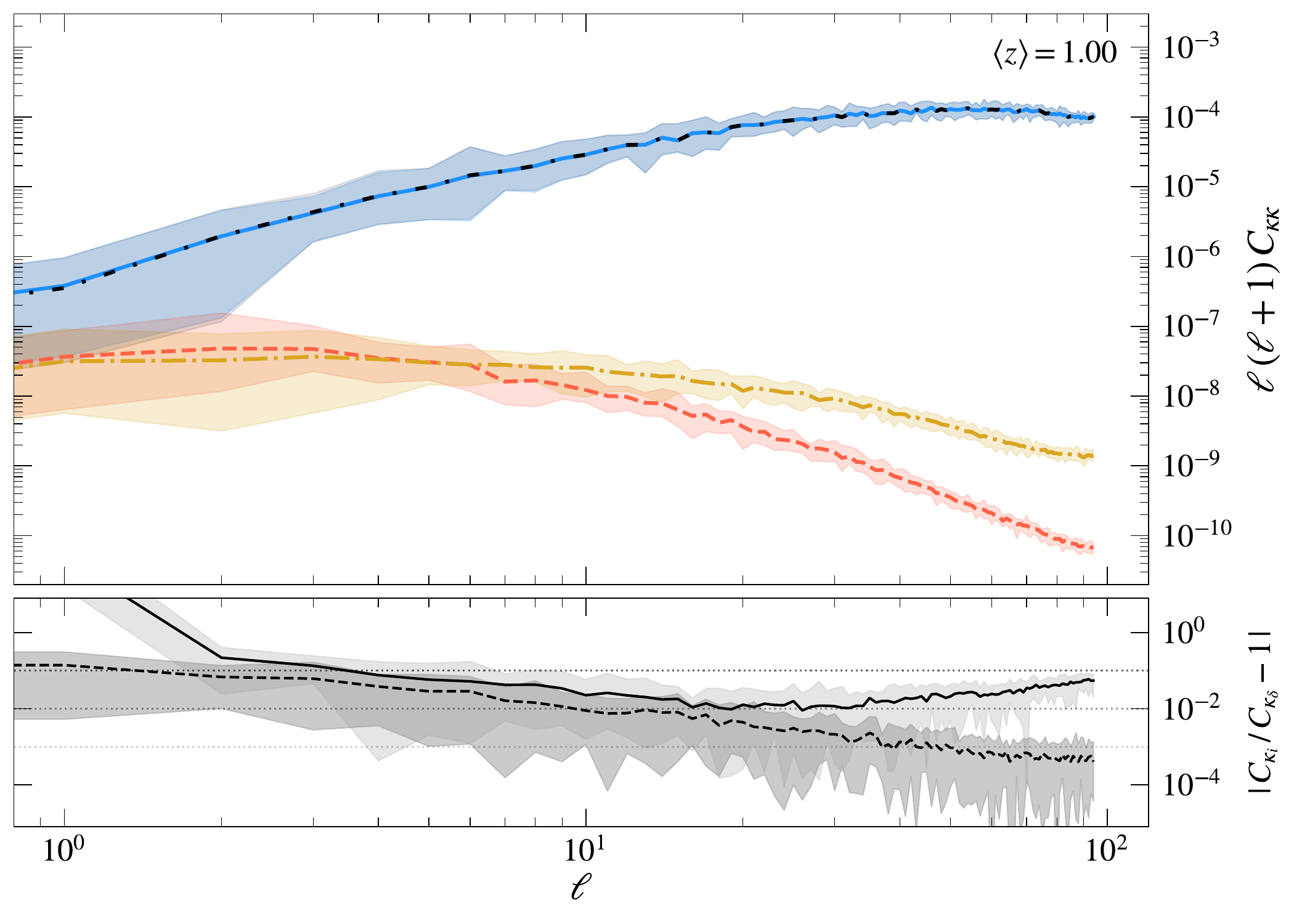}
    \end{minipage}
    \caption{Angular power spectra of $\kappa_{{\rm d}^2\phi}$ (yellow dot-dashed curve), $\kappa_{{\rm d}\phi}$ (red dashed curve), their sum $\kappa_g$ (black sparse dot-dashed curve), and $\kappa_\delta$ (blue solid curve). Left panel shows $\langle z \rangle=0.5$ and right panel is $\langle z \rangle=1.0$. Curves show the mean power spectra across all 20 observers and the shaded regions of the same colour indicate the spread from minimum to maximum across observers. Bottom panels for each redshift show the difference in power spectra between $\kappa$ and $\kappa_\delta$ (solid curve and light shaded region) and between $\kappa_\delta$ and $\kappa_g$ (dashed curve and dark shaded region).}
    \label{fig:kappa_xtra}
\end{figure*}
Figure~\ref{fig:kappa_xtra} shows the angular power spectra for each of the additional contributions in \eqref{eqs:kextras} at two redshift slices. In both panels, solid blue curves show the $\kappa_\delta$ power spectrum presented in the main text, red dashed curves show $\kappa_{{\rm d}\phi}$ in \eqref{eq:kdphi}, yellow dot-dashed curves show $\kappa_{{\rm d}^2\phi}$ in \eqref{eq:kd2phi}, and black sparse dot-dashed curves show the sum $\kappa_g$ in \eqref{eq:kappas_neglected}. Curves are the mean power spectra across all 20 observers and shaded regions show the range of minimum to maximum across observers. Bottom panels for each plot shows the absolute value of the relative difference in power spectra between $\kappa_\delta$ and $\kappa$ (solid curve and light shaded region) and the difference between $\kappa_\delta$ and $\kappa_g$ (dashed curve and dark shaded region). The solid curves here are the same as the dotted curves in the lower panels of Figure~\ref{fig:Cl_comp_all}. Horizontal dotted lines in the bottom panels indicate 10\%, 1\%, and 0.1\% differences, for reference. All power spectra shown here exhibit numerical convergence up to $\ell=70$ for $\langle z \rangle = 0.5$ and to $\ell_{\rm max}$ for $\langle z \rangle = 1.0$, which is the maximum $\ell$ for each respective plot.

Since the difference between $\kappa_\delta$ alone and when we include additional, neglected terms (dashed curves) is below the difference between the nonlinear convergence and $\kappa_\delta$ (solid curves) we can conclude that in these simulations the neglected terms can, in fact, be neglected safely.

\subsection{Validity of the Poisson equation}\label{appx:del2phi}

The first term on the right hand side of \eqref{eq:kappas_neglected} has already been simplified from its original form (the first term on the right hand side of \eqref{eq:nabla_perp_phi}), which is 
\begin{equation}
    \kappa_{\nabla^2\phi} \equiv \int^{\chi_s}_0 {\rm d}\chi \frac{\chi (\chi_s - \chi)}{\chi_s} \nabla^2\phi, \label{eq:kdel2phi}
\end{equation}
where we have $\kappa_{\nabla^2\phi}=\kappa_\delta$ so long as the Poisson equation holds. 
In this Appendix we will calculate $\kappa_{\nabla^2\phi}$ directly and compare it to $\kappa_\delta$ to ensure this relation holds. We extract $\nabla^2\phi$ from the simulation via the 3--Ricci scalar, which for the linearly-perturbed metric \eqref{eq:pertmetric} is
\begin{equation}\label{eq:3Rdel2phi}
    \mathcal{R} = \frac{4}{a^2}\nabla^2 \phi.
\end{equation}
The \texttt{mescaline} post-processing analysis calculates the Ricci scalar directly from the spatial metric $\gamma_{ij}$ (i.e., it does not assume the form \eqref{eq:pertmetric}) and interpolates it to each point along the geodesic. We use this output to obtain $\nabla^2\phi$ along each line of sight to subsequently calculate \eqref{eq:kdel2phi}. 

\begin{figure}
    \centering
    \includegraphics[width=\columnwidth]{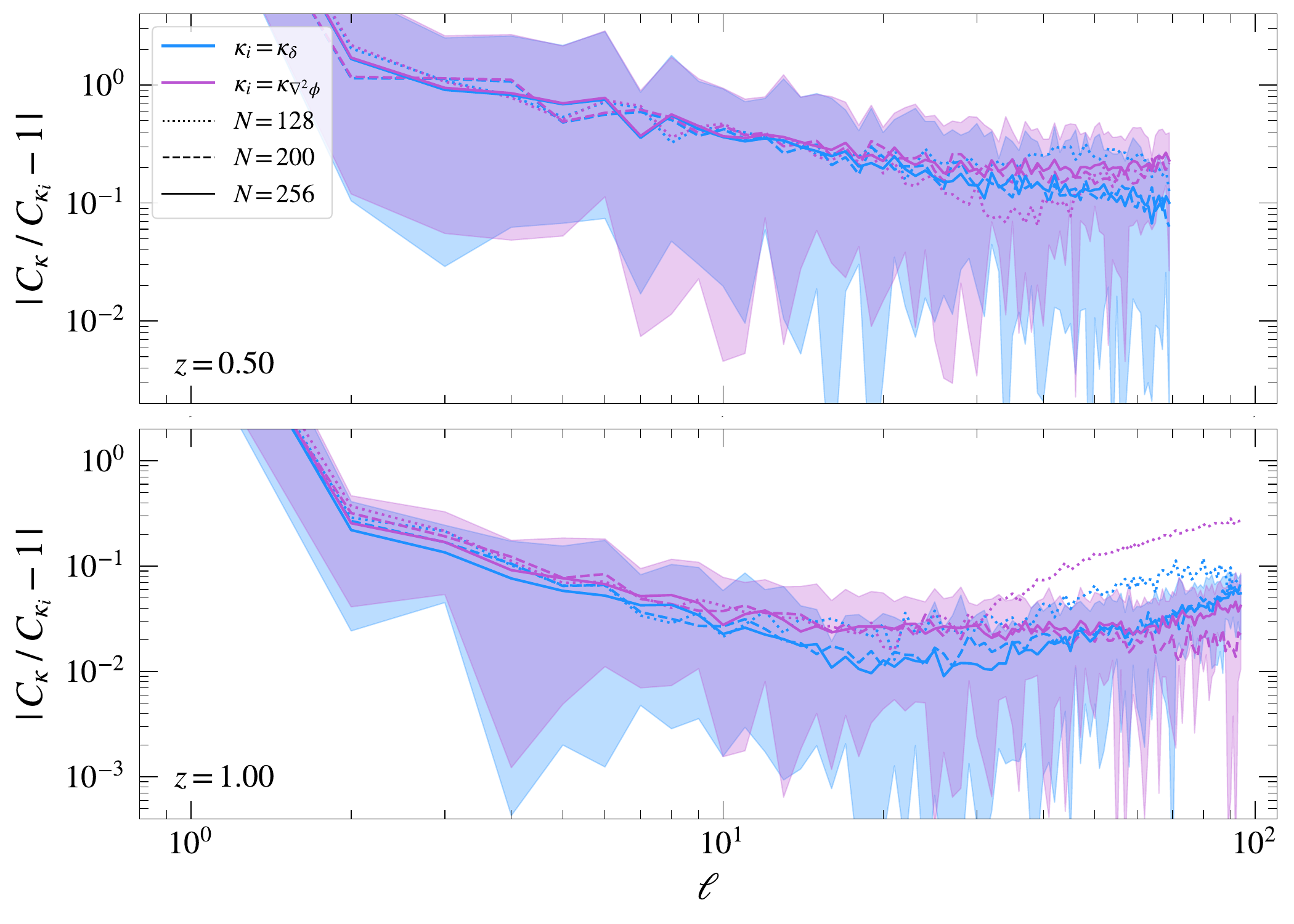}
    \caption{Absolute relative difference in power spectra between the nonlinear convergence $\kappa$ and the approximations $\kappa_\delta$ (blue curves and shaded regions) and $\kappa_{\nabla^2\phi}$ (purple curves and shaded regions). Different line styles indicate different numerical resolutions, as indicated in the legend. Top panel shows $\langle z \rangle = 0.5$ and bottom panel shows $\langle z \rangle = 1.0$. Curves are the mean across all 20 observers and shaded regions show the spread from minimum to maximum absolute difference across all observers.}
    \label{fig:Cldiff_del2phi}
\end{figure}
Figure~\ref{fig:Cldiff_del2phi} shows the difference in the angular power spectra between the nonlinear convergence, $\kappa$, and the two expressions for the linearised convergence from gravitational lensing. Blue curves and shaded regions show the difference for $\kappa_\delta$ and purple curves and shaded regions for $\kappa_{\nabla^2\phi}$. The top panel shows the light cone slice with $\langle z \rangle = 0.5$ and the bottom panel shows $\langle z \rangle = 1.0$. In both panels, curves show the mean difference over 20 observers and shaded regions show the full spread from minimum to maximum. Different line styles indicate calculations in different resolution simulations; $N=128$ (dotted), $N=200$ (dashed), and $N=256$ (solid) so that we can also directly see for which scales we have numerical convergence. We exclude the dipole from the plots since the difference reaches amplitudes of $\sim 10^4$--$10^5$ and we are more interested in focusing on the higher multipoles since the dipole is dominated by peculiar motion. 

Overall, we find little difference between $\kappa_\delta$ and $\kappa_{\nabla^2\phi}$ when considering the full spread across observers. At $\langle z \rangle=0.5$ (top panel of Figure~\ref{fig:Cldiff_del2phi}), we do see the two signals begin to diverge from each other at $\ell \gtrsim$~40--50 and this is robust to numerical resolution (i.e. solid and dashed curves agree). As in the main text, we cut the top panel at $\ell=70$ since $\kappa_\delta$ is not numerically converged beyond this point. 
At higher redshift of $\langle z \rangle = 1.0$ (bottom panel), we note a divergence in the average differences at $10 \lesssim \ell \lesssim 40$ which is robust to resolution. At $\ell\sim 50$, the difference for $\kappa_{\nabla^2\phi}$ is no longer robust to resolution (i.e. solid and dashed curves begin to diverge) for this redshift. 

Our conclusion from this test is that the Poisson equation is not broken sufficiently to explain the difference we find here. However, the difference in $\kappa_\delta$ and $\kappa_{\nabla^2\phi}$ at small angular scales and low redshift is interesting and motivates further investigation with higher-resolution ray tracing maps.

\end{document}